\documentclass[12pt]{article}
\pdfoutput=1
\usepackage[utf8]{inputenc}

\usepackage{here}
\usepackage{graphicx}
\usepackage[numbers,sort&compress]{natbib}
\usepackage[margin=25mm]{geometry} 
\usepackage[T1]{fontenc}
\usepackage{amssymb,amsmath,amsthm,amsfonts}
\usepackage{enumerate,mathrsfs}
\usepackage{xcolor}
\usepackage{cancel}
\usepackage[squaren]{SIunits}
\usepackage{mathtools}
\usepackage{prettyref}
\usepackage{blkarray}
\usepackage{adjustbox}
\usepackage{array,booktabs,tabularx,fancybox}
\usepackage{comment}
\usepackage[normalem]{ulem}
\usepackage{afterpage}
\usepackage{tikz}
\usepackage{tikz-feynhand}
\usepackage{nicematrix}
\usepackage{kbordermatrix}
\usepackage{nicematrix}
\usetikzlibrary{decorations.pathmorphing}
\usetikzlibrary{decorations.markings}
\usetikzlibrary{calc}
\usetikzlibrary{fit, shapes, tikzmark}
\usepackage{listings}
\usepackage{authblk}

\usepackage{hyperref}
\usepackage{cleveref}
\definecolor{gh-fg}{HTML}{000000}
\definecolor{gh-bg}{gray}{0.95}
\definecolor{gh-comment}{HTML}{007ACC}
\definecolor{gh-keyword}{HTML}{E60000}
\definecolor{gh-func}{HTML}{630330}
\definecolor{gh-string}{HTML}{999999} 
\definecolor{gh-number}{HTML}{000000}
\definecolor{gh-linenum}{HTML}{666666}
\definecolor{gh-frame}{HTML}{BBBBBB}
\definecolor{gh-type}{HTML}{007ACC}

\usepackage{setspace}
\lstdefinelanguage{Singular}{
  morekeywords={
    LIB,ring,ideal,poly,int,intvec,module,matrix,
    list,def,setring,Weyl,option,print,
    proc,return,if,else,for,while,break,continue
  },
  sensitive=true,
  morecomment=[l]{//},
  morecomment=[s]{/*}{*/},
  morestring=[b]",
}

\lstdefinestyle{githubSingular}{
  language=Singular,
  backgroundcolor=\color{gh-bg},
  basicstyle={\small\setstretch{1}\def\fvm@Scale{0.85}\ttfamily\selectfont\color{gh-fg}},
  commentstyle=\itshape\color{gh-comment},
  keywordstyle=\bfseries\color{gh-fg},
  stringstyle=\color{gh-string},
  identifierstyle=\color{gh-fg},
  numbers=left,
  numberstyle={\scriptsize\sffamily\color{gh-linenum}},
  numbersep=8pt,
  frame=lines,
  rulecolor=\color{gh-frame},
  tabsize=4,
  showstringspaces=false,
  breaklines=false,
  numbers=none,
  columns=fixed,
  basewidth=0.5em,
  aboveskip=1em,
  belowskip=0pt,
  emph=[1]{ring,ideal,poly,int,intvec,module,matrix,list},
  emphstyle=[1]\bfseries\color{gh-type},
  emph=[2]{holonomicRank,restrictionIdeal,Weyl},
  emphstyle=[2]\bfseries\color{gh-func},
  literate=
    *{0}{{{\color{gh-number}0}}}1 {1}{{{\color{gh-number}1}}}1
     {2}{{{\color{gh-number}2}}}1 {3}{{{\color{gh-number}3}}}1
     {4}{{{\color{gh-number}4}}}1 {5}{{{\color{gh-number}5}}}1
     {6}{{{\color{gh-number}6}}}1 {7}{{{\color{gh-number}7}}}1
     {8}{{{\color{gh-number}8}}}1 {9}{{{\color{gh-number}9}}}1
     {.0}{{{\color{gh-number}.0}}}2 {.1}{{{\color{gh-number}.1}}}2
     {.2}{{{\color{gh-number}.2}}}2 {.3}{{{\color{gh-number}.3}}}2
     {.4}{{{\color{gh-number}.4}}}2 {.5}{{{\color{gh-number}.5}}}2
     {.6}{{{\color{gh-number}.6}}}2 {.7}{{{\color{gh-number}.7}}}2
     {.8}{{{\color{gh-number}.8}}}2 {.9}{{{\color{gh-number}.9}}}2
}

\usepackage{lipsum}
\newtheorem{theorem}{Theorem}[section]
\newtheorem{proposition}[theorem]{Proposition}
\newtheorem{lemma}[theorem]{Lemma}
\newtheorem{example}[theorem]{Example}

\newtheorem{corollary}[theorem]{Corollary}
\numberwithin{equation}{section}
\newtheorem{definition}[theorem]{Definition}
\newtheorem*{namedtheorem}{Theorem 6.1}

\definecolor{cyan1}{HTML}{37cdaa}
\definecolor{blue1}{HTML}{5d7ac4}
\definecolor{red1}{HTML}{d0482a}
\definecolor{purple1}{HTML}{845ea8}
\definecolor{orange1}{HTML}{e07229}
\definecolor{yellow1}{HTML}{edcb52}
\definecolor{green}{HTML}{244819}
\definecolor{red}{HTML}{921818}
\definecolor{purple}{HTML}{53047A}
\definecolor{yellow}{HTML}{f4e097}
\definecolor{zoomFuchsia}{HTML}{CB29BA}
\definecolor{zoomYellow1}{HTML}{FFE5B0}
\definecolor{zoomYellow2}{HTML}{FFD47B}
\definecolor{zoomYellow3}{HTML}{D08D00}
\definecolor{zoomPurple1}{HTML}{DBC9FF}
\definecolor{zoomPurple2}{HTML}{C2A4FF}
\definecolor{zoomPurple3}{HTML}{5300F9}
\definecolor{zoomGreen1}{HTML}{C2EBCE}
\definecolor{zoomGreen2}{HTML}{99DDAD}
\definecolor{zoomGreen3}{HTML}{339951}
\definecolor{zoomBlue1}{HTML}{ABD1FF}
\definecolor{zoomBlue2}{HTML}{73B2FF}
\definecolor{zoomBlue3}{HTML}{005AC8}
\definecolor{zoomRed1}{HTML}{FBC2C6}
\definecolor{zoomRed2}{HTML}{F899A0}
\definecolor{zoomRed3}{HTML}{D70F1E}
\definecolor{zoomOrange1}{HTML}{FFCEA7}
\definecolor{zoomOrange2}{HTML}{FFC291}
\definecolor{zoomOrange3}{HTML}{C15500}
\definecolor{gr}{gray}{0.7}
\definecolor{gr1}{gray}{0.7}

\newcommand{\lrbrk}[1]{\left(#1\right)}
\newcommand{\bigbrk}[1]{\bigl(#1\bigr)}
\newcommand{\Bigbrk}[1]{\Bigl(#1\Bigr)}

\newcommand{\brc}[1]{\{#1\}}

\newcommand{\vev}[1]{\langle #1\rangle}

\DeclarePairedDelimiterX\braket[2]{\langle}{\rangle}{#1 \>\delimsize\vert\> #2}

\newcommand{\defas}{:=}

\newcommand{\ii}{\sqrt{-1}}

\def\CC{{\mathbb C}}

\def\pd#1{\partial_{#1}}
\def\ZZ{{\mathbb Z}}

\makeatletter
\newcommand*\bigcdot{\mathpalette\bigcdot@{1}}
\newcommand*\smtimes{\mathpalette\smtimes@{.7}}
\newcommand*\bigcdot@[2]{\mathbin{\vcenter{\hbox{\scalebox{#2}{$\m@th#1\bullet$}}}}}
\newcommand*\smtimes@[2]{\mathbin{\vcenter{\hbox{\scalebox{#2}{$\m@th#1\times$}}}}}
\makeatother

\providecommand{\filename}[1]{}
\renewcommand{\filename}[1]{\texttt{#1}}

\title{Banana D-modules}

\author{Giacomo Brunello, Wojciech Flieger, Yoshiaki Goto,
\vspace{-1em}
\\Saiei-Jaeyeong Matsubara-Heo, Nobuki Takayama}

\date{}
\begin{document}

\maketitle

\begin{abstract}
We study the holonomic $D$-module whose solutions are generic-mass banana Feynman integrals with complex propagator powers and spacetime dimension. We call this the banana $D$-module.
Two properties of the banana $D$-module are studied: its holonomic rank and the irreducibility of the monodromy representation.
This is achieved by relating the banana $D$-module to Lauricella $F_{C}$ hypergeometric systems.
Further, we expand the generic-mass banana Feynman integral and the associated discontinuity integral in terms of series solutions.
Finally, we analyze the Pfaffian system arising from the banana $D$-module and construct an explicit $d\log$ representation in the equal-mass specialization.
\end{abstract}
\section{Introduction}
Scattering amplitudes are theoretical tools describing particle interactions, used to build physical observables, providing a direct bridge between quantum field theory and collider experiments.
Precision predictions are typically obtained within perturbation theory, where scattering amplitudes are expanded in powers of the relevant coupling constants.
The coefficients of this expansion are expressed in terms of Feynman integrals.
Feynman integrals are intricate multivalued functions defined through integral representations.
It was recognised early on that uncovering their mathematical structure not only facilitates their evaluation but can also reveal deeper properties of the underlying quantum field theory.
An early manifestation of this perspective was the use of differential equations by Regge and collaborators to investigate the analytic properties of Feynman integrals~\cite{Regge:1965}.
Building on this viewpoint, Regge further suggested that, owing to their ramification structure, Feynman integrals should be regarded as generalisations of classical hypergeometric functions~\cite{Golubeva_1976}.
In the same period, the monodromy of Feynman integrals was studied in~\cite{Ponzano:1969tk}, and their holonomic character was established by Kashiwara and Kawai~\cite{Kashiwara:1977yy}.
Decades later, together with integration-by-parts identities~\cite{Chetyrkin:1981qh}, the method of differential equations was rediscovered and intensively developed, becoming the state of the art for the computation of Feynman integrals~\cite{Kotikov:1990kg,Remiddi:1997ny,Henn:2013pwa}.
On the mathematical side, Gelfand, Kapranov, and Zelevinsky developed the theory of $A$-hypergeometric systems, commonly called GKZ systems, and of the associated hypergeometric functions~\cite{Gelfand:1990}.
Feynman integrals can be regarded as special cases of Euler integrals, or more generally of twisted period integrals~\cite{Mastrolia:2018uzb} (see also Ref.~\cite{Frellesvig:2019uqt}), and thus as specialisations of solutions of GKZ systems~\cite{delaCruz:2019skx,Klausen:2019hrg,Feng:2019bdx}.
This viewpoint underlies Macaulay method approaches to integral relations~\cite{Chestnov:2022alh}, as well as $D$-module methods for Feynman integrals~\cite{Bitoun:2017nre,Henn:2023tbo,Chestnov:2025whi}.
Their singular loci are governed by the Landau equations~\cite{Landau:1959fi} and have recently been described with tools from computational algebraic geometry~\cite{Fevola:2023kaw}.

Among the various families of Feynman integrals, banana integrals provide a particularly useful laboratory for exploring general mathematical structures.

They are sufficiently structured to admit a systematic study at arbitrary loop order while already exhibiting geometries that extend beyond those underlying multiple polylogarithms.
As a result, they have become an important testing ground for techniques based on differential equations, periods, algebraic geometry, and special functions.
Starting from the two-loop sunrise integral~\cite{Berends:1993ee,Caffo:1998du,Laporta:2004rb,Muller-Stach:2011qkg,Adams:2013nia}, banana integrals have been studied at arbitrary loop order from the viewpoints of maximal cuts, Picard--Fuchs operators and Calabi--Yau geometry~\cite{Bloch:2016izu,Primo:2017ipr,Klemm:2019dbm,Bonisch:2020qmm,Mishnyakov:2023wpd,Vanhove:2026fth}, and have been computed explicitly at three and four loops~\cite{Pogel:2022yat,Pogel:2022ken,Duhr:2025kkq,Pogel:2025bca}.
The application of Mellin–Barnes methods has revealed a connection between generic mass banana integrals and multivariable hypergeometric functions~\cite{Berends:1993ee, Kalmykov:2016lxx}~(see also~\cite{Boos:1990rg}).

In~\cite{Flieger:2025bho}, the second author introduced a $D$-module whose solutions are banana Feynman integrals with an arbitrary spacetime dimension and all the propagator powers set to one.
For this $D$-module to be the correct one, the holonomic rank (i.e., the dimension of the space of holomorphic solutions) should be $2^n-1$, which is the number of master integrals of the generic mass banana Feynman diagram~\cite{Kalmykov:2016lxx,Bitoun:2017nre,Flieger:2025bho} (see also~\cite{Lee:2013hzt}).

Let us summarise our results below.
The main object studied in this paper is the banana $D$-module $M(\alpha)$ introduced in Definition \ref{def:banana D-module}.
Here, $\alpha=(\alpha_0,\dots,\alpha_n)$ is a complex parameter vector. 
The banana $D$-module $M(\alpha)$ is a generalization of the $D$-module introduced in~\cite{Flieger:2025bho} which can be recovered for $\alpha_0=\frac{d}{2}$ and $\alpha_1=\cdots=\alpha_n=2-\frac{d}{2}$.
More generally, the generic mass banana Feynman integral with its propagator powers regarded as a part of the parameters $\alpha$ is a solution of the banana $D$-module.
The $D$-module $M(\alpha)$ is also a special example of the matroid hypergeometric system introduced in~\cite{matsubara2026principal}.
The following theorem answers affirmatively~\cite[Conjecture 1]{Flieger:2025bho}.

\begin{theorem}
    The holonomic rank of $M(\alpha)$ is $2^n-1$.   
\end{theorem}
We prove the theorem by establishing matching lower and upper bounds. For the lower bound, after choosing $j$ and introducing the ratios $z_i = \frac{x_i}{x_j}$, we identify a subsystem of $M(\alpha)$ with the Lauricella $F_{C}$ system~\cite[Chapter 3]{koornwinder2020multivariable}.
Its local solution space has dimension $2^{n}$. 
An additional operator imposes another independent relation, leaving $2^{n}-1$ solutions.
The upper bound is obtained independently from the structure of the standard monomials of $M(\alpha)$.
We construct an explicit basis of $2^{n}-1$ local solutions $F_J(z,\alpha)$, identified by proper subsets $J\subsetneq\{0,\dots,n\}\setminus\{j\}$, in terms of Lauricella $F_{C}$ functions.  

By specializing parameters to Feynman integral values $\alpha_0=\frac{d}{2}$, and $\alpha_i=\nu_i-\frac{d}{2}+1$ for $i=1,\dots,n$, where $\nu_{i}$ are propagator powers, and using the Mellin-Barnes representation of Feynman integrals we show in Theorem \ref{thm:expansion} that the banana integrals when $|x_i|<\!\!<|x_0|$ for any $i=1,\dots,n$ can be expressed in terms of $F_J(z,\nu)$.
Although such an expansion previously appeared in~\cite{Bonisch:2021yfw}, we include a short, independent proof of Theorem \ref{thm:expansion} to keep our exposition self-contained.

Another result concerns the monodromy representation of the banana $D$-module $M(\alpha)$.
Based on the monodromy analysis of Lauricella $F_{C}$ functions in~\cite{Goto-FC-monodromy, GotoMatsumoto-FC}~(see also~\cite{hattori-takayama-2014}; for the analogous irreducibility criterion of $A$-hypergeometric systems see~\cite{SchulzeWalther:2012}) we prove the following theorem.
\begin{namedtheorem}
  If the condition 
  \begin{align}
    \sum_{j\in J} \alpha_j \notin \ZZ, \quad \emptyset \neq J \subset \brc{0,1,\dots,n}
  \end{align}
  holds, the monodromy representation of $M(\alpha)$ is irreducible.
\end{namedtheorem}
\noindent
As a byproduct of the monodromy analysis of $M(\alpha)$ we obtain Theorem \ref{th: ban_combination} which expands discontinuity integrals in terms of $F_J(z;\nu)$; for related discussions of cuts, discontinuities and the monodromy group of Feynman integrals see~\cite{Cutkosky:1960sp,Bourjaily:2020wvq}.

Finally, we study the Pfaffian system associated with $M(\alpha)$ and show how one can construct such a Pfaffian system from the ideal $J_n(\alpha)$ using the Differential-Difference Weyl algebra~\cite{OharaTakayama:2009}. We show in Theorem \ref{thm:impossible} based on the geometry of the singular locus of $M(\alpha)$ that when $n \geq 2$ the monodromy representation of the Pfaffian system in the logarithmic connection form ($d \log$ form) is reducible when the integrability condition is satisfied.  
For the equal mass case $x_1=x_2 = \hdots = x_n$, we construct in Theorem \ref{thm:em_main} the Pfaffian system in the logarithmic form explicitly and its basis is given by
\begin{equation}
    J_{1}=K_{L}\bullet f \,,
    \qquad
    J_{j+1}=x^{\,j-c}\;\partial_{x}^{\,j-1}\bullet f \,,
    \qquad j=1,\dots,L \,,
\end{equation}
where $f$ is the equal mass banana Feynman integral and $K_{L}$ is the differential operator introduced in~\eqref{eq: KL operator}, $L=n-1$ and $c=\bigl\lceil \frac{n}{2}\bigr\rceil$.
Theorem~\ref{thm:em_main} produces a $d\log$ form, not a canonical one~\cite{Henn:2013pwa}.
The canonical ($\varepsilon$-factorised) form is instead reached in~\cite{Pogel:2022yat,Pogel:2022ken,Pogel:2022vat,Duhr:2025tdf,Pogel:2025bca,Duhr:2025kkq}
by admitting transcendental letters built from the periods of the Calabi--Yau $(L-1)$-fold. 
The paper is organised as follows. Section \ref{sec: feynman integrals} reviews Feynman integrals and the banana family. Sections \ref{sec:the_lower_bound_of_the_holonomic_rank}  and \ref{sec:the_upper_bound} establish the lower and upper bounds for the holonomic rank. Section \ref{sec:Expansion of the banana} derives the Lauricella expansion of the banana integrals, and Section \ref{sec:monodromy} studies monodromy and discontinuities. Section \ref{sec:pfaffian system} develops the associated Pfaffian systems and the obstruction to logarithmic form. Section \ref{sec:equalmass} treats the equal-mass specialisation with integer propagator powers.
\section{Feynman integrals}\label{sec: feynman integrals}
\begin{figure}[H]
    \centering
    \includegraphics[width=0.7\linewidth]{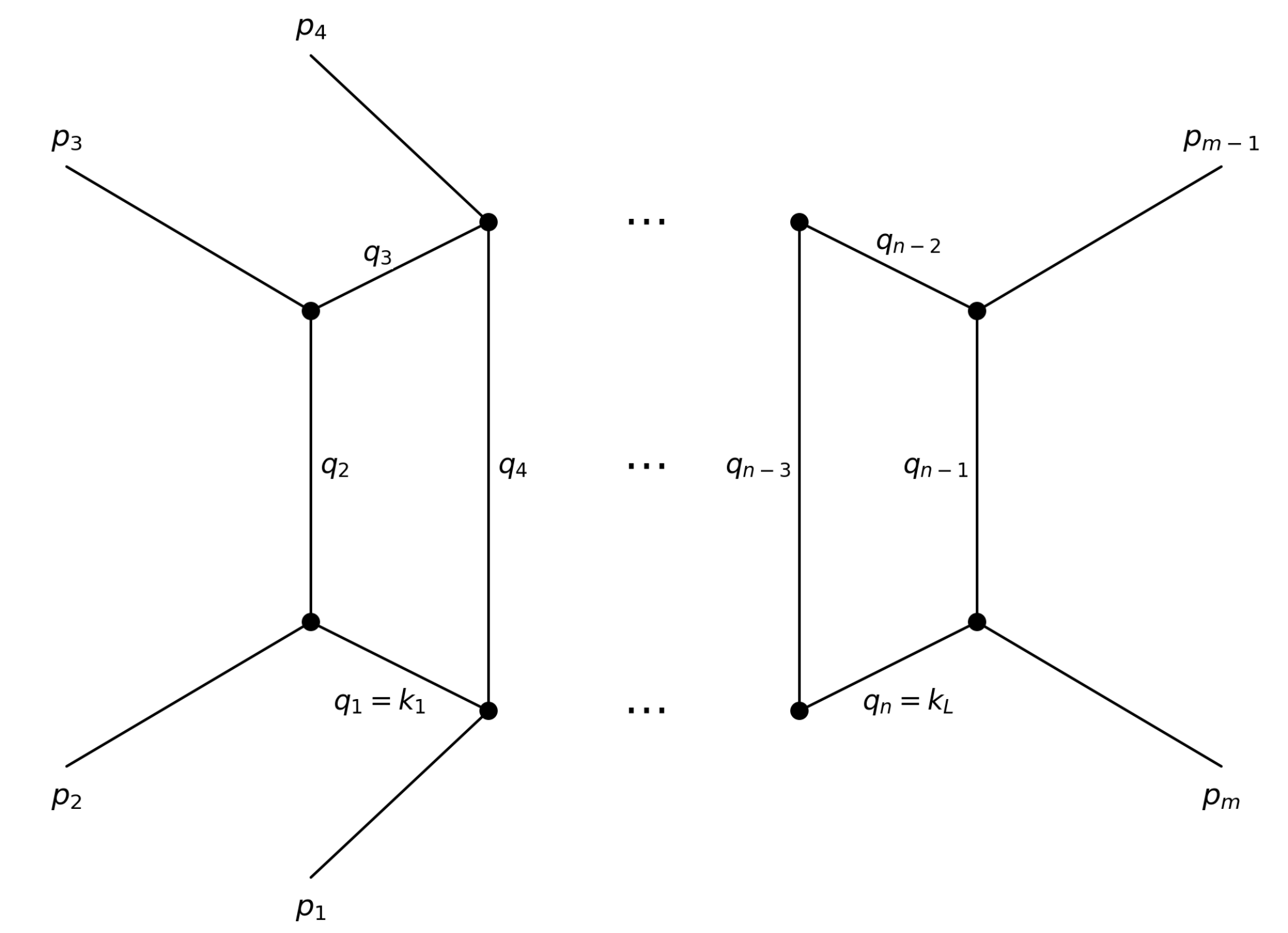}
    \caption{$L$-loop Feynman diagram}
    \label{fig:feynman_diagram}
\end{figure}

The computation of scattering amplitudes in perturbative quantum field theory is characterised by the appearance of multi-loop multi-leg integrals known as Feynman integrals.
The Feynman integral associated with the Feynman diagram in Fig.~\ref{fig:feynman_diagram}, with $L$ loop momenta $\{k_i\}_{i=1}^L$, and $m$ external momenta $\{p_j\}_{j=1}^m$, is given in momentum space representation by
\begin{align} \label{eq:momentum_space_integral}
I_{\nu_{1}, \hdots , \nu_{n}} := e^{\varepsilon L\gamma_E}\int \prod_{r=1}^{L}\frac{d^{d}k_{r}}{\ii \pi^{\frac{d}{2}} }\prod_{j=1}^{n} \frac{1}{(q_{j}^{2}-m_{j}^{2})^{\nu_{j}}} \,,    
\end{align}
where $q_{j} = \sum_{i=1}^{L} \lambda_{ji} k_{i} + \sum_{r=1}^{m-1} \sigma_{jr} p_{r}$ are internal momenta, being the sum of the loop momenta and external momenta, and $m_j$ are the masses associated to each internal propagator. Propagator powers are denoted by $\nu_{j} \in \ZZ$, $d\in\mathbb{C}$ is the regularised dimension with $\varepsilon$ being the dimensional regularisation parameter such that $d+2\varepsilon\in\ZZ$ is the true dimension, $L$ is the number of loops, and $\gamma_E$ is Euler's constant.
A sector associated to $I_{\nu_{1}, \hdots , \nu_{n}}$ is characterized by the set of indices $i$ with $\nu_{i}>0$ in the tuple $(\nu_{1}, \hdots , \nu_{n})$.
Thus, the integrals where all $\nu_{j}>0$ are called top sector integrals and if one or more $\nu_{j} < 1$, we say the integral belongs to a sub-sector.
One can also talk about discontinuities of Feynman integrals when crossing the singular locus also known as cut of Feynman integrals\footnote{On a Feynman diagram this corresponds to cutting some of the internal edges which reflects these propagators going ``on-shell'', i.e., $q_j^2=m_j^2$.}. 
Here, a singular locus of $I_{\nu_1,\dots,\nu_n}$ is an analytic variety $S$ in the space of kinematic variables where it fails to be holomorphic.
Note that such a locus is necessarily codimension one~\cite[Chapter 2, Theorem 5B]{WhitneyAnalyticVarieties}.
Let $\rho$ be a simple loop which goes around an irreducible component $S$ of the singular locus.
The discontinuity integral is defined by
\begin{equation}
{\rm Disc}_\rho (I_{\nu_1,\dots,\nu_n}):=\rho_*I_{\nu_1,\dots,\nu_n}-I_{\nu_1,\dots,\nu_n}\,,    
\end{equation}
where $\rho_*I_{\nu_1,\dots,\nu_n}$ is the analytic continuation of $I_{\nu_1,\dots,\nu_n}$ along $\rho$.
To be more precise, the discontinuity integral should be regarded as an element of the quotient space of the space of multivalued functions ramified along $S$ by the space of holomorphic functions.
When the irreducible component comes from the Landau equation~\cite[Eq.s (5)-(7)]{Landau:1959fi} for some cuts, a formula for these discontinuities when all $\nu_{j}=1$ was provided by Cutkosky~\cite{Cutkosky:1960sp}.

In this work, we will focus on the Feynman parametric representation:
\begin{equation}\label{eq:Feynman parametrization}
I_{\nu_1,\dots,\nu_n}=(-1)^{|\nu|}\frac{e^{\varepsilon L\gamma_E}\Gamma(|\nu|-\frac{Ld}{2})}{\Gamma(\nu_1)\cdots\Gamma(\nu_n)}\int_{\mathbb{R}^{n-1}_+}\frac{\mathcal{U}(a)^\mu}{\mathcal{F}(a)^{\mu'}}a^{\nu-{\bf 1}}\Omega(a)\,,
\end{equation}
where $a=(a_1,\dots,a_n)$ are Schwinger parameters, $\nu=(\nu_1,\dots,\nu_n)$ are propagator powers, $\mu=|\nu|-\frac{(L+1)d}{2}$, $\mu'=|\nu|-\frac{Ld}{2}$, $\mathcal{U},\mathcal{F}$ are the first and the second Symanzik polynomials, $\Omega(a)=\sum_{i=1}^n(-1)^{i-1}a_id a_1\cdots\widehat{d a_i}\cdots d a_n$.

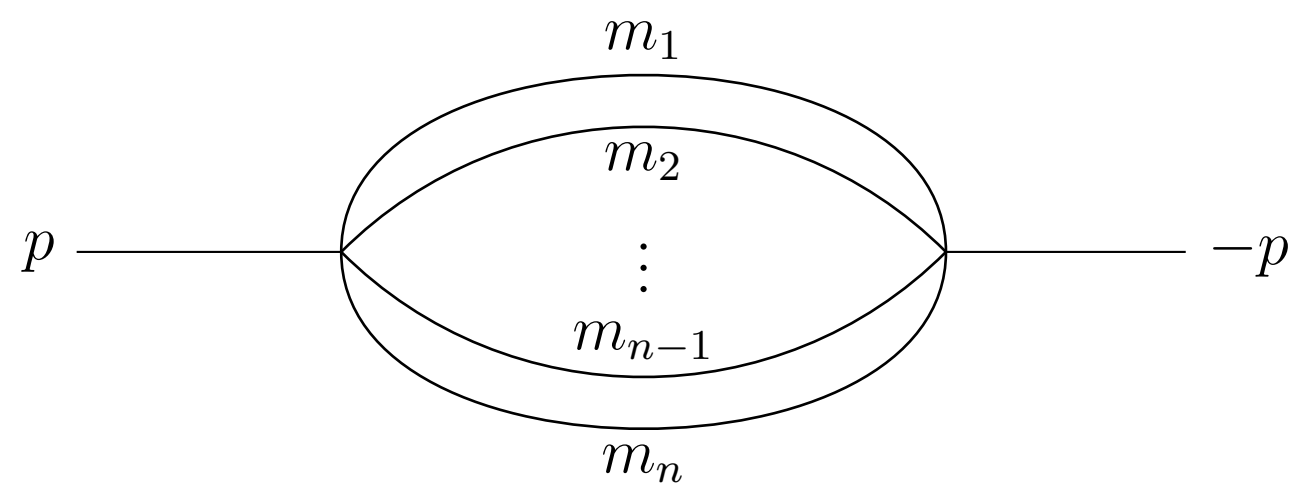
\begin{figure}
 \begin{center}
\begin{tikzpicture}[scale=2]
    \begin{feynhand}
    \vertex (a) at (0,0);
    \vertex (b) at (2,0);
    \vertex [particle] (p1) at (-1,0){$p$};
    \vertex [particle] (p2) at (3,0){$-p$};
    \propag (p1) to (a);
    \propag (p2) to (b);
    \propag [plain] (a) to [out=90, in=90](b);
    \propag [plain] (a) to [out=-90, in=-90](b);
    \propag [plain] (a) to [out=-45, in=-135](b);
    \propag [plain] (a) to [out=45, in=135](b);
    \end{feynhand}
    \node at (1,.7){$m_{1}$};
    \node at (1,.3){$m_{2}$};
    \node at (1,0){$\vdots$};
    \node at (1,-.3){$m_{n-1}$};
    \node at (1,-.7){$m_n$};
\end{tikzpicture}
\end{center}
    \caption{Banana diagram}
    \label{fig:banana}
\end{figure}

The main focus of this work is the $L$-loop banana diagram with different masses, shown in Fig.~\ref{fig:banana}. This integral topology has one external momentum $p$ and $n$-internal legs, each associated with a different mass $m_i$, $i\in \{1,\ldots,n\}$.
Fixing $L=(n-1)$, the Symanzik polynomials are given by
\begin{equation}\label{eq:symanzik_poly}
\mathcal{U}=a_1\cdots a_n\sum_{i=1}^n a_i^{-1},\quad \quad \mathcal{F}=\left(\sum_{i=1}^nx_i a_i\right)\mathcal{U}-x_0 a_1\cdots a_n \,,
\end{equation}
where we denote the kinematic variables as $x_{0} = p^{2}$ and $x_{i} = m_{i}^{2}$ for $i\in \{1,\ldots , n\}$.

Let $D$ be the Weyl algebra $D=\mathbb{C}\langle x_0,\cdots,x_n,\partial_0,\dots,\partial_n\rangle$, where $\partial_i$ represents the partial derivative with respect to $x_i$.
\begin{definition}\label{def:banana D-module}
Let $\alpha=(\alpha_0,\dots,\alpha_n)\in\CC^{n+1}$ be a complex vector and set
\begin{equation}\label{eq:constraint}
    \beta=n-\sum_{i=0}^n\alpha_i\,.
\end{equation}
Let $J_n(\alpha)$ be the left ideal of $D$ generated by the following operators:
\begin{eqnarray}
{\cal A}_E(\beta) &:=& \sum_{i=0}^n x_i \pd{i} - \beta \label{eq:A_E beta}\\
{\cal A}_{ij}(\alpha) &:=& x_i \pd{i}^2+\alpha_i \pd{i} - 
                         (x_j \pd{j}^2+\alpha_j \pd{j}), \ 0 \leq i \not= j \leq n \,,\label{eq:A_ij alpha} \\
\mathcal{A}_S &:=& \sum_{i=0}^{n} \partial_{0}\cdots \hat{\partial}_{i} \cdots \partial_{n} \,.\label{eq:1.3}
\end{eqnarray}
We set $M(\alpha):=D/J_n(\alpha)$ and call it the banana $D$-module.
\end{definition}

\section{The lower bound of the holonomic rank} \label{sec:the_lower_bound_of_the_holonomic_rank}

After the change of integration variables $a_i\mapsto a_i^{-1}$ in~\eqref{eq:Feynman parametrization} for the banana diagram, we obtain that
\begin{equation}\label{eq:matroid integral}
I_{\nu_1,\dots,\nu_n}\propto(-1)^{n-1}\int_{\mathbb{R}_+^{n-1}} a^{-1-\nu+\frac{d}{2}{\bf 1}}\left( \sum_{i=1}^n a_i\right)^{-\frac{d}{2}}\left(\frac{x_0}{-\sum_{i=1}^n a_i}+\sum_{i=1}^n\frac{x_i}{a_i}\right)^{-\mu'}\Omega(a)\,.
\end{equation}
By~\cite[Proposition 5.4]{matsubara2026principal}, we obtain that $I_{\nu_1,\dots,\nu_n}$ is a solution to the $D$-module $M(\alpha)$ where 
\begin{equation}\label{eq:propagators and alpha}
\alpha_0=\frac{d}{2},\quad\quad\alpha_i=\nu_i-\frac{d}{2}+1\ \ (i=1,\dots,n)\,.    
\end{equation}
In the rest of this section, we prove the following assertion:
\begin{theorem}\label{thm:lower bound}
    The holonomic rank of the module $M(\alpha)$ is at least $2^n-1$.
\end{theorem}

The proof is based on an explicit construction of independent solutions to $M(\alpha)$ and a limiting procedure.
Recall that Lauricella's $F_C$ function is given by the following series:
\begin{equation}
    F_C\left(\substack{a,b\\
        c_1,\dots,c_n};
        x_1,\dots,x_n\right)
        =
        \sum_{m=(m_1,\dots,m_n)\in\ZZ^n_{\geq 0}}\frac{(a)_{|m|}(b)_{|m|}}{(c_1)_{m_1}\cdots(c_n)_{m_n}m_1!\cdots m_n!}x_1^{m_1}\cdots x_n^{m_n}\,.
\end{equation}
Here, we put $|m|:=m_1+\cdots+m_n$.
Let $J_n(\alpha,\beta)\subset D$ be the ideal generated by~\eqref{eq:A_E beta} and~\eqref{eq:A_ij alpha}.
In the following theorem, we do not need to impose the constraint~\eqref{eq:constraint}.

\begin{theorem}\label{thm:Jn-FC}
Fix any $0\leq j\leq n$ and set $z_i = \frac{x_i}{x_j}$,  for any $i \not= j$ and $z_j=x_j$.
Then, a holomorphic function $f$ is annihilated by the left ideal $J_n(\alpha,\beta)$ if and only if it is annihilated by
\begin{eqnarray}
&&\theta_j - \beta  \label{eq:theta-beta}\\
&&\theta_i (\theta_i + \alpha_i-1) -z_i 
  \left( \sum_{k=0, k \not= j}^n \theta_k - \beta \right)
  \left( \sum_{k=0, k \not= j}^n \theta_k - \beta - \alpha_j + 1\right)\,.
  \label{eq:F_C}
\end{eqnarray}
Here $i$ runs over $\{0, 1, \ldots, n\} \setminus \{ j\}$ and $\theta_k = z_k \frac{\partial}{\partial z_k}$.
\end{theorem}

\begin{proof}
In this proof, $\theta_{x_i}$ denotes $x_i \pd{i}$ unless otherwise stated in this proof.
Since $x_i^2 \pd{i}^2 = \theta_{x_i} (\theta_{x_i}-1)$,
we have
$$
 x_i x_j {\cal A}_{ij}(\alpha,\beta)
= x_j \theta_{x_i} (\theta_{x_i}+\alpha_i-1) - x_i \theta_{x_j} (\theta_{x_j}+\alpha_j-1)\,.
$$
By the relation $\theta_{x_j} \equiv -\sum_{k=0, k\not=j}^n \theta_{x_k} + \beta$,
it can be rewritten as
$$
\theta_{x_i} (\theta_{x_i}+\alpha_i-1) 
 - \frac{x_i}{x_j} 
   \left( \sum_{k=0, k\not=j}^n \theta_{x_k} - \beta   \right)
   \left( \sum_{k=0, k\not=j}^n \theta_{x_k} - \beta -\alpha_j +1  \right)\,.
$$
By the coordinate transformation $z_i = \frac{x_i}{x_j}$, $i \not= j$, $z_j=x_j$,
we have
$$ x_i \frac{\partial}{\partial x_i} = z_i \frac{\partial}{\partial z_i}, \quad i \not=j \ \ \ \text{and}\ \ \  x_j \frac{\partial}{\partial x_j} = -\sum_{k=0, k\not=j}^n z_k \frac{\partial}{\partial z_k}+z_j\frac{\partial}{\partial z_j}\,.
$$
Thus, we obtain the result. 
\end{proof}

The system~\eqref{eq:F_C} is Lauricella system $F_C$
whose local solutions at the origin are expressed in terms of 
the Lauricella function $F_C$ as
$$ z_j^\beta F_C\left(\substack{-\beta, -\beta-\alpha_j+1\\ {\tilde \alpha}}; {\tilde z}\right)\,,$$
where 
$\tilde \alpha = (\alpha_0, \ldots, \alpha_{j-1},\alpha_{j+1}, \ldots, \alpha_n)$
and 
$\tilde z = (z_0, \ldots, z_{j-1},z_{j+1}, \ldots, z_n)$.
For any subset $J\subset \{0,1,\dots,n\}\setminus\{ j\}$, the following functions span the solution space if none of $\alpha_i$ is an integer for any $i\in\{0,\dots,n\}\setminus\{j\}$:
\begin{equation}\label{eq:Lauricella}
z_j^\beta\prod_{i\in J}z_i^{1-\alpha_i} F_C\left(\substack{-\beta+|J|-\sum_{k\in J}\alpha_k, -\beta-\alpha_j+1+|J|-\sum_{k\in J}\alpha_k\\ (2-\alpha_i)_{i\in J},(\alpha_i)_{i\notin J,i\neq j}}; {\tilde z}\right)\,.
\end{equation}

The rank, characteristic variety, and reducibility of the system $F_C$ 
(for any parameter value) are given in~\cite{hattori-takayama-2014}.

\begin{theorem}\label{thm:solution basis}
Assume that none of $\alpha_i$ is an integer for any $i\neq j$ and that $\beta=n-|\alpha|$.
Then, the series~\eqref{eq:Lauricella} is annihilated by the operator~\eqref{eq:1.3} if and only if $J\subsetneq\{0,\dots,n\}\setminus\{j\}$.
\end{theorem}

\begin{proof}
    We expand the series~\eqref{eq:Lauricella} as
    \begin{align}
        &z_j^\beta\prod_{i\in J}z_i^{1-\alpha_i} F_C\left(\substack{-\beta+|J|-\sum_{k\in J}\alpha_k, -\beta-\alpha_j+1+|J|-\sum_{k\in J}\alpha_k\\ (2-\alpha_i)_{i\in J},(\alpha_i)_{i\notin J,i\neq j}}; {\tilde z}\right)\nonumber\\
        =&\sum_{m=(m_i)_{i\neq j},m_i\geq 0}\frac{(-\beta+|J|-\sum_{k\in J}\alpha_k)_{|m|}(-\beta-\alpha_j+1+|J|-\sum_{k\in J}\alpha_k)_{|m|}}{\prod_{i\in J}(2-\alpha_i)_{m_i}\prod_{i\notin J,i\neq j}(\alpha_i)_{m_i}m!}\times\nonumber\\
        &x_j^{\beta-|m|-|J|+\sum_{k\in J}\alpha_k}\prod_{i\in J}x_i^{m_i+1-\alpha_i}\prod_{i\notin J, i\neq j}x_i^{m_i}\,.\label{eq: Lauricella expanded}
    \end{align}
    If $J\subsetneq\{0,\dots,n\}\setminus\{ j\}$, we may assume from the continuity that
    $$
    -\beta+|J|-\sum_{k\in J}\alpha_k,\ \  -\beta-\alpha_j+1+|J|-\sum_{k\in J}\alpha_k\notin\ZZ_{>0}\,.
    $$ 
    Therefore, we may assume that the summation index $m$ runs over $\ZZ^n$ in~\eqref{eq: Lauricella expanded} by the relation $m_i!=\Gamma(1+m_i)$.
    Therefore, it follows that 
    \begin{equation}\label{eq:5.7}
        \begin{split}
            &\prod_{i=0,i\neq j}^n\pd{i}\bullet z_j^\beta\prod_{i\in J}z_i^{1-\alpha_i} F_C\left(\substack{-\beta+|J|-\sum_{k\in J}\alpha_k, -\beta-\alpha_j+1+|J|-\sum_{k\in J}\alpha_k\\ (2-\alpha_i)_{i\in J},(\alpha_i)_{i\notin J,i\neq j}}; {\tilde z}\right)\\
        =&\sum_{m=(m_i)_{i\neq j},m_i\in\ZZ}\frac{(-\beta+|J|-\sum_{k\in J}\alpha_k)_{|m|+n}(-\beta-\alpha_j+1+|J|-\sum_{k\in J}\alpha_k)_{|m|+n}}{\prod_{i\in J}(2-\alpha_i)_{m_i}\prod_{i\notin J,i\neq j}(\alpha_i)_{m_i}\prod_{i\in J}m_i!\prod_{i\notin J,i\neq j}m_i!}\times\\
        &\frac{1}{\prod_{i\notin J,i\neq j}(\alpha_i+m_i)\prod_{i\in J}(m_i+1)}x_j^{\beta-|m|-n-|J|+\sum_{k\in J}\alpha_k}\prod_{i\in J}x_i^{m_i+1-\alpha_i}\prod_{i\notin J, i\neq j}x_i^{m_i}\,.
        \end{split}
    \end{equation}
    Note that $-\beta+|J|-\sum_{k\in J}\alpha_k=\alpha_j$ and $-\beta-\alpha_j+1+|J|-\sum_{k\in J}\alpha_k=1$ if $J=\{0,\dots,n\}\setminus\{ j\}$.
 Consider an index $k\notin J$ and $k\neq j$.
    Then,
    \begin{equation}\label{eq:5.8}
        \begin{split}
            &\prod_{i=0,i\neq k}^n\pd{i}\bullet z_j^\beta\prod_{i\in J}z_i^{1-\alpha_i} F_C\left(\substack{-\beta+|J|-\sum_{k\in J}\alpha_k, -\beta-\alpha_j+1+|J|-\sum_{k\in J}\alpha_k\\ (2-\alpha_i)_{i\in J},(\alpha_i)_{i\notin J,i\neq j}}; {\tilde z}\right)\\
        =&-\sum_{m=(m_i)_{i\neq j},m_i\in\ZZ}\frac{(-\beta+|J|-\sum_{k\in J}\alpha_k)_{|m|+n}(-\beta-\alpha_j+1+|J|-\sum_{k\in J}\alpha_k)_{|m|+n-1}}{\prod_{i\in J}(2-\alpha_i)_{m_i}\prod_{i\notin J,i\neq j}(\alpha_i)_{m_i}\prod_{i\in J}m_i!\prod_{i\notin J,i\neq j}m_i!}\times\\
        &\frac{1}{\prod_{i\notin J,i\neq j,k}(\alpha_i+m_i)\prod_{i\in J}(m_i+1)}x_j^{\beta-|m|-n-|J|+\sum_{k\in J}\alpha_k}\prod_{i\in J}x_i^{m_i+1-\alpha_i}\prod_{i\notin J, i\neq j}x_i^{m_i}\,.
        \end{split}
    \end{equation}
    Similarly, for an index $k\in J$, we have the following identities.
    \begin{equation}\label{eq:5.9}
        \begin{split}
            &\prod_{i=0,i\neq k}^n\pd{i}\bullet z_j^\beta\prod_{i\in J}z_i^{1-\alpha_i} F_C\left(\substack{-\beta+|J|-\sum_{k\in J}\alpha_k, -\beta-\alpha_j+1+|J|-\sum_{k\in J}\alpha_k\\ (2-\alpha_i)_{i\in J},(\alpha_i)_{i\notin J,i\neq j}}; {\tilde z}\right)\\
        =&-\sum_{m=(m_i)_{i\neq j},m_i\in\ZZ}\frac{(-\beta+|J|-\sum_{k\in J}\alpha_k)_{|m|+n}(-\beta-\alpha_j+1+|J|-\sum_{k\in J}\alpha_k)_{|m|+n-1}}{\prod_{i\in J}(2-\alpha_i)_{m_i}\prod_{i\notin J,i\neq j}(\alpha_i)_{m_i}\prod_{i\in J}m_i!\prod_{i\notin J,i\neq j}m_i!}\times\\
        &\frac{1}{\prod_{i\notin J,i\neq j}(\alpha_i+m_i)\prod_{i\in J,i\neq k}(m_i+1)}x_j^{\beta-|m|-n-|J|+\sum_{k\in J}\alpha_k}\prod_{i\in J}x_i^{m_i+1-\alpha_i}\prod_{i\notin J, i\neq j}x_i^{m_i}\,.
        \end{split}
    \end{equation}
    Summing up the equations~\eqref{eq:5.7},~\eqref{eq:5.8} and~\eqref{eq:5.9}, we obtain that
    \begin{equation}
        \begin{split}
 &\left(\prod_{i=0}^n\pd{i}\sum_{i=0}^n\pd{i}^{-1}\right)\bullet z_j^\beta\prod_{i\in J}z_i^{1-\alpha_i} F_C\left(\substack{-\beta+|J|-\sum_{k\in J}\alpha_k, -\beta-\alpha_j+1+|J|-\sum_{k\in J}\alpha_k\\ (2-\alpha_i)_{i\in J},(\alpha_i)_{i\notin J,i\neq j}}; {\tilde z}\right)\\
=&\sum_{m=(m_i)_{i\neq j},m_i\in\ZZ}\frac{(-\beta+|J|-\sum_{k\in J}\alpha_k)_{|m|+n}(-\beta-\alpha_j+1+|J|-\sum_{k\in J}\alpha_k)_{|m|+n-1}}{\prod_{i\in J}(2-\alpha_i)_{m_i}\prod_{i\notin J,i\neq j}(\alpha_i)_{m_i}\prod_{i\in J}m_i!\prod_{i\notin J,i\neq j}m_i!}\times\\
        &\frac{C}{\prod_{i\notin J,i\neq j}(\alpha_i+m_i)\prod_{i\in J}(m_i+1)}x_j^{\beta-|m|-n-|J|+\sum_{k\in J}\alpha_k}\prod_{i\in J}x_i^{m_i+1-\alpha_i}\prod_{i\notin J, i\neq j}x_i^{m_i}\,,
        \end{split}
    \end{equation}
    where
    \begin{equation}
        C=-\beta-\alpha_j+|m|+n+|J|-\sum_{k\in J}\alpha_k-\sum_{k\notin J,k\neq j}\alpha_k-|J|-|m|=-\beta+n-|\alpha|=0\,.
    \end{equation}
\end{proof}

\begin{example}
    In~\eqref{eq:Lauricella}, we set $n=2$, $j=0$, $J=\{2\}$, and use the substitution~\eqref{eq:propagators and alpha} with $\nu_1=0$.
    Then, the function~\eqref{eq:Lauricella} becomes
    \begin{equation}\label{eq:tadpole integral}
        x_2^{\frac{d}{2}-\nu_2}F_4\left(\substack{0,-\frac{d}{2}+1\\ \frac{d}{2}-\nu_2+1,1-\frac{d}{2}}; \frac{x_1}{x_0},\frac{x_2}{x_0}\right)=x_2^{\frac{d}{2}-\nu_2}\,.
    \end{equation}
    This is the tadpole integral~\cite[Equation (41)]{abreu2022sagex}.
\end{example}

Theorem \ref{thm:solution basis} shows that the holonomic rank of $M(\alpha)$ is at least $2^{n}-1$ when none of $\alpha_i$ is an integer.
Let us conclude this section with a proof of Theorem \ref{thm:lower bound}.
In fact, the following argument is well-known and reproduced in many places in the literature.

\vspace{1em}

\noindent
(Proof of Theorem \ref{thm:lower bound})

It is enough to prove the theorem when some of the parameters $\alpha_i$ are integral.
Let $v\in\CC^{n+1}$ be a generic vector so that no entry of  $\alpha(\varepsilon):=\alpha+\varepsilon v$ is integral.
Let $F_J(x;\alpha)$ denote the function~\eqref{eq:Lauricella} with the constraint~\eqref{eq:constraint}.
Then, the functions $\{ F_J(x;\alpha(\varepsilon))\}_{J\subsetneq \{0,\dots, n\}\setminus\{ j\}}$ satisfy the assumption of~\cite[Proposition 2.21]{oshima1980eigenspaces}.
Thus, one can find Laurent series $\{ c_J^{(k)}(\varepsilon)\}_{J\subsetneq \{0,\dots, n\}\setminus\{ j\}}$ $(k=1,\dots,2^n-1)$ such that the limit $\lim_{\varepsilon\to 0}\sum_{J\subsetneq \{0,\dots, n\}\setminus\{ j\}}c_J^{(k)}(\varepsilon)F_J(x;\alpha(\varepsilon))$ exists for any $k=1,\dots,2^n-1$ and they are linearly independent.
Thus, we have constructed $2^n-1$ linearly independent solutions to $M(\alpha)$.
\qed

\section{The upper bound }  \label{sec:the_upper_bound}
In this section, we first prove the following theorem.

\begin{theorem} \label{thm:upper_bound}
    The upper bound of the holonomic rank of 
$M(\alpha)$ is $2^n-1$. 
\end{theorem}

To prove this theorem, we use the following theorem about $D$-modules.
Recall that for a differential operator $P=\sum_{\alpha,\beta}c_{\alpha,\beta}x^\alpha\partial^\beta\in D$, its order ${\rm ord}(P)$ is defined as ${\rm max}\{ |\beta|\mid \exists\alpha, c_{\alpha,\beta}\neq 0\}$ and the principal symbol $\sigma(P)$ is defined as $\sigma(P):=\sum_{\alpha,\beta={\rm ord}(P)}c_{\alpha,\beta}x^\alpha\xi^\beta\in \CC[x_0,\ldots, x_n,\xi_0, \ldots, \xi_n]=:\CC[x,\xi]$.
For a left ideal $I\subset D$, its characteristic ideal $I_{\rm char}\subset\CC[x,\xi]$ is defined as the ideal generated by the elements $\sigma(P)$ for $P\in I$.

\begin{theorem} {\rm (see, e.g.,~\cite{MR1299957},~\cite[\S 1.4]{SST} for algorithmic evaluation)} \label{thm:rank_is_multiplicity}
Let $I \subset D$ be a holonomic ideal.
The holonomic rank of the $D$-module $D/I$ is equal to the multiplicity
\begin{equation}
\begin{split}
 {\rm dim}_\CC\, \CC[x,\xi]/I_{\rm char}\otimes_{\CC[x]}(\CC[x]/\langle x_0-a_0,\dots,x_n-a_n\rangle)&={\rm dim}_{\CC(x)}\, \CC[x,\xi]/I_{\rm char}\otimes_{\CC[x]}\CC(x)\\
 &={\rm dim}_{F}\, \CC[x,\xi]/I_{\rm char}\otimes_{\CC[x]}F\,,
 \end{split}
\end{equation}
where $a=(a_0,\dots,a_n)$ is any point in $\CC^{n+1}$ out of the singular locus of $D/I$ and $\CC(x)\subset F$ is any field extension.
\end{theorem}

The set of the principal symbols of $M(\alpha)$ contains
$ \sum_{i=0}^n x_i \xi_i$,
$ -x_0 \xi_0^2 + x_i \xi_i^2$, $i=1, \ldots, n$.
Consider the ideal generated by these elements.
We will evaluate the multiplicity of the ideal 
with respect to the $\xi$ variables for generic values of $x$.
Since these elements belong to the characteristic ideal, the multiplicity of the ideal generated by these elements gives an upper bound of the holonomic rank
of $D/J_n(\alpha,\beta)$.

By a change of variables $z_i = \sqrt{x_i} \xi_i$,
the ideal is generated by
$ \sum_{i=0}^n \sqrt{x_i} z_i$,
$ -z_0^2 + z_i^2$, $i=1, \ldots, n$.
This transformation of variables makes the ideal easier to consider.
We put $c_i=\sqrt{x_i}$ and will evaluate the multiplicity.
This amounts to considering a field extension $\CC(x)\subset \CC(c)=:K$ given by $x_i\mapsto c_i^2$.

\begin{proposition}\label{prop:basis 2n}
Put
\begin{equation} \label{eq:h}
h=c_0 z_0 + c_1 z_1 + \cdots + c_n z_n\,,
\end{equation}
and let $I$ be the ideal generated by
$ -z_0^2 + z_i^2$, $i=1, \ldots, n$
in $K[z_0, \ldots, z_n]$.
The $K$-vector space
$K[z_0, z_1, \ldots, z_n]/(I,h)$
is a finite dimensional vector space with 
$2^n$ basis vectors
$$  z_1^{e_1} \cdots z_n^{e_n},
 \ e_i \in \{0, 1\}\,.
$$
\end{proposition}
\begin{proof}
The set 
$ \{ z_0^2- \underline{z_i^2}\,|\, i=1, \ldots, n\}$
is a Gr\"obner basis of $I$ with respect to the
lexicographic order $z_n \succ \cdots \succ z_1 \succ z_0$
where $\underline{\ }$ denotes the leading monomial.
Put 
$ R=K[z_0, z_1, \ldots, z_n]/I $
and 
$R_k$ denotes the subvector space of degree $k$
elements of $R$.
\begin{eqnarray*}
R_0 &=& K \cdot 1 \\
R_1 &=& K z_0 + K z_1 + \cdots + K z_n \\
R_2 &=& K z_0^2 + K z_0 z_1 + \cdots + K z_0 z_n 
                  + K z_1 z_2 + \cdots + K z_{n-1}z_n \\
    & &  \vdots \\
R_{n} &=& K z_0^n + K z_0^{n-1} z_1 + \cdots + K z_0^{n-1} z_n + \cdots + z_1 z_2 \cdots z_n \\
    & & \vdots\,. 
\end{eqnarray*}
Thus, 
${\rm dim}_K R_k = \sum_{i=0}^k {n \choose i}$ 
 when $k \leq n$
and
${\rm dim}_K R_k = 2^n$ when $n \leq k$.

Consider the degree one homogeneous polynomial
$h$ defined in (\ref{eq:h}).
We have
$$ K[z_0, z_1, \ldots, z_n]/(I,h)
 = R/hR
\ \ \text{and}\ \  (R/hR)_{k+1} = R_{k+1}/hR_{k}\,. $$
The linear homogeneous polynomial $h$ defines
a linear map $H_k$ from $R_{k}$ to $R_{k+1}$.
When $k \geq n$, the matrix representation of 
linear map $H_k$ is a $2^n \times 2^n$ matrix.
If
$c_1 = \cdots = c_n = 0$, it is a diagonal matrix and the determinant
is equal to $c_0^{2^n}$.
Therefore, the rank of the linear map $H_k$
is $2^n$, 
which means that
\begin{equation}\label{eq:H_k}
H_k : R_k \longrightarrow R_{k+1}    
\end{equation}
is an isomorphism and then
$R_{k+1}/h R_k = 0$.

Let us consider the case $k < n$.
We consider the special case $n=3$ and $k=1$.
The following is the matrix $H_1$ that represents
the map from $R_1$ to $R_2$.
\[
H_1 = \begin{pNiceArray}{cccc|ccc}[first-row, first-col]
    & z_0^2 & z_0z_1 & z_0z_2 & z_0z_3 & z_1z_2 & z_1z_3 & z_2z_3 \\
z_0 & c_0   & c_1    & c_2    & c_3    & 0      & 0      & 0      \\
z_1 & c_1   & c_0    & 0      & 0      & c_2    & c_3    & 0      \\
z_2 & c_2   & 0      & c_0    & 0      & c_1    & 0      & c_3    \\
z_3 & c_3   & 0      & 0      & c_0    & 0      & c_1    & c_2
\end{pNiceArray}\,.
\]
Here, we use the relation $z_i^2 \equiv z_0^2$,
$i=1,2,3$.
When $c_0 \not= 0$ and $c_1=c_2=c_3=0$,
the $4 \times 4$ submatrix in the left corner
of the matrix $H_1$ above is non-singular.
Thus, $ H_1 : R_1 \rightarrow R_2$
is the rank $4$ map.
It follows that ${\rm dim}_K (R_2/hR_1) = (4+3)-4=3$.
Moreover, we can see that $R_2/hR_1$ is spanned by
$ z_1z_2, z_1 z_3, z_2 z_3$.
We can show $H_k$ is of maximal rank for other cases
analogously for general $n$ and $k<n$.
\end{proof}

Based on the above preparations,
let us prove the following theorem which implies Theorem \ref{thm:upper_bound}.
To state it, we put 
\begin{equation} \label{eq:S}
S=\sum_{i=0}^n c_i \prod_{j \not= i} z_j\,. 
\end{equation}

\begin{theorem} \label{prop:rank_upper_bound}
The $K$-vector space
$K[z_0, z_1, \ldots, z_n]/(I,h,S)$ is a finite dimensional vector space 
with a basis consisting of $2^n-1$ elements
$$
  z_1^{e_1} \cdots z_n^{e_n}, \quad e_i \in \{0,1\}, \ 
 (e_1, \ldots, e_n) \not= (1, \ldots, 1)\,.
$$
\end{theorem}

\begin{proof}
The matrix representation of $H_{n-1}$ defined in~\eqref{eq:H_k} by a row vector representation
is
$ \left( P | v \right)$
where $P$ is a $p \times p$ ($p=2^n-1$) square matrix whose diagonal entries are $c_0$ 
and $v^T=(0, \ldots, 0, c_n, \ldots, c_1)$.
The last column vector $v$ is indexed by the monomial
$\prod_{i=1}^n z_i$ in $R_n$.
For example, when $n=3$ we have
{\footnotesize
$$
H_2=\begin{pNiceArray}{ccccccc|c}
 {c}_{0}&  {c}_{1}&  {c}_{2}&  {c}_{3}& 0& 0& 0& 0 \\
 {c}_{1}&  {c}_{0}& 0& 0&  {c}_{2}&  {c}_{3}& 0& 0 \\
 {c}_{2}& 0&  {c}_{0}& 0&  {c}_{1}& 0&  {c}_{3}& 0 \\
 {c}_{3}& 0& 0&  {c}_{0}& 0&  {c}_{1}&  {c}_{2}& 0 \\
0&  {c}_{2}&  {c}_{1}& 0&  {c}_{0}& 0& 0&  {c}_{3} \\
0&  {c}_{3}& 0&  {c}_{1}& 0&  {c}_{0}& 0&  {c}_{2} \\
0& 0&  {c}_{3}&  {c}_{2}& 0& 0&  {c}_{0}&  {c}_{1} \\
\end{pNiceArray}\,, 
$$
}
where the basis of $R_2$ is
$r_2=( {z}_{0}^{ 2} ,  {z}_{0}  {z}_{1},  {z}_{0}  {z}_{2},  {z}_{0}  {z}_{3},  {z}_{1}  {z}_{2},  {z}_{1}  {z}_{3},  {z}_{2}  {z}_{3})$
and the basis of $R_3$ is
$ r_3=(  {z}_{0}^{ 3} ,   {z}_{0}^{ 2}   {z}_{1},   {z}_{0}^{ 2}   {z}_{2},   {z}_{0}^{ 2}   {z}_{3},   {z}_{0}  {z}_{1}  {z}_{2},   {z}_{0}  {z}_{1}  {z}_{3},   {z}_{0}  {z}_{2}  {z}_{3},   {z}_{1}  {z}_{2}  {z}_{3})$.
Assume $c_0 \not=0$ and fix its value.
Let $r_n$ be the basis of $R_n$.
Each element of $Hr_n^T$ belongs to $h R_{n-1}$.
Therefore, each element of $P^{-1} H r_n^T$ also belongs to $h R_{n-1}$.
Note that $P^{-1}H = (E_{p}|P^{-1}v)$ where $E_{p}$ is the $p \times p$
identity matrix.
Hence, the elements of the basis of $R_n$ but the last element
are equal to $- P^{-1}v \prod_{i=1}^n z_i$ modulo $(I,h)$.
In our example,
$(z_0^3, z_0^2 z_1, \ldots, z_0 z_2 z_3 )^T \equiv - P^{-1}v \prod_{i=1}^3 z_i$.
Since the determinant of $P$ is $c_0^p$ when $c_i=0$, ($i=1, \ldots, n$),
elements of the vector $P^{-1} v$ do not have a pole in $c_1,\dots,c_n$.
In particular, $z_0 \prod_{j \not=0, i} z_j$ for $1\leq i\leq n$ can be expressed in terms of
$z_1 \cdots z_n$ 
in $K[z_0, \ldots, z_n]/(I,h)$
and the coefficient of it does not have a pole in $c_1,\dots,c_n$.
The polynomial $S$ can be written as 
$$
c_0 z_1 \cdots z_n + z_0 \sum_{i=1}^n c_i \prod_{j\not=0, i} z_j\,,
$$
and it is congruent to 
$$ 
 \left(c_0 - o(c_1,\dots,c_n)\right) z_1 \cdots z_n 
$$
modulo $(I,h)$.
The proof is completed by the above discussion.
\end{proof}

\noindent
(Proof of Theorem \ref{thm:upper_bound})
The principal symbol of the operator ${\cal A}_S$ is
$$
 \sum_{i=0}^n \prod_{j \not= i} \xi_j\,.
$$
By the change of coordinates $z_j=c_j\xi_j$,
the symbol is equal to 
$$ 
  \sum_{i=0}^n \prod_{j \not= i} z_j/c_j
= \frac{1}{\prod_{j=0}^n c_j} \left( \sum_{i=0}^n c_i \prod_{j \not= i} z_j \right)= \frac{1}{\prod_{j=0}^n c_j} S\,.
$$
Thus, the theorem follows from Theorem \ref{thm:rank_is_multiplicity} and Theorem \ref{prop:rank_upper_bound}.

\section{Expansion of the banana integral in terms of Lauricella series}\label{sec:Expansion of the banana}
Since the Feynman integral~\eqref{eq:Feynman parametrization} is a solution to $M(\alpha)$, it is expanded into a linear combination of a basis of its solutions.
In this section, we explicitly give a formula of such an expansion.
Under the substitution~\eqref{eq:propagators and alpha}, the solutions~\eqref{eq:Lauricella} with $j=0$ take the following form:
\begin{equation}\label{eq:Lauricella banana}
F_J(x;\nu):=x_0^{\frac{(n-1)d}{2}-|\nu|}\prod_{i\in J}\left(\frac{x_i}{x_0}\right)^{\frac{d}{2}-\nu_i} F_C\left(\substack{-\frac{(n-1-|J|)d}{2}+|\nu_{\bar{J}}|,-\frac{(n-|J|)d}{2}+|\nu_{\bar{J}}|+1\\ (\frac{d}{2}-\nu_i+1)_{i\in J},(\nu_i+1-\frac{d}{2})_{i\notin J,i\neq 0}}; \frac{x_1}{x_0},\dots,\frac{x_n}{x_0}\right)\,,
\end{equation}
where $\bar{J}=\{1,\dots,n\}\setminus J$ and $\nu_{\bar{J}}:=(\nu_j)_{j\in \bar{J}}$.
We follow Vanhove's derivation of the Mellin-Barnes integral of the banana Feynman integral~\cite{Vanhove:2026fth}.
A similar derivation in dimensional regularization was presented in~\cite{Bonisch:2021yfw}.
Instead of Feynman's parametric representation~\eqref{eq:Feynman parametrization}, we use the exponential Schwinger representation:
\begin{equation}\label{eq:Schwinger representation}
    I_{\nu_1,\dots,\nu_n}=(-1)^{|\nu|}\frac{e^{\varepsilon L\gamma_E}}{\Gamma(\nu_1)\cdots\Gamma(\nu_n)}\int_{\mathbb{R}_+^n}\mathcal{U}(a)^{-\frac{d}{2}}\exp\left\{ -\frac{\mathcal{F}(a)}{\mathcal{U}(a)}\right\}a^{\nu}\frac{d a}{a}\,.
\end{equation}

It is easy to see that $I_{\nu_1,\dots,\nu_n}$ is proportional to the following integral:
\begin{equation}
    \tilde{I}_{\nu_1,\dots,\nu_n} :=\int_{\mathbb{R}_+^n}\left(\sum_{i=1}^n a_i^{-1}\right)^{-\frac{d}{2}}\exp\left\{-\sum_{i=1}^nx_i a_i+\frac{x_0}{\sum_{i=1}^n a^{-1}_i}\right\} a^{\nu-\frac{d}{2}{\bf 1}}\frac{d a}{a}\,.
\end{equation}
The exact relation is $I_{\nu_1,\dots,\nu_n}=(-1)^{|\nu|}\frac{e^{\varepsilon L\gamma_E}}{\Gamma(\nu_1)\cdots\Gamma(\nu_n)}\tilde{I}_{\nu_1,\dots,\nu_n}.$

\begin{theorem}
    The following relation is true:
    \begin{equation}\label{eq:Mellin-Barnes representation}
    \begin{split}
        \tilde{I}_{\nu_1,\dots,\nu_n}=&\frac{1}{(2\pi\ii)^n}\int_{(\ii\mathbb{R}-\varepsilon)^n}\\
        &\frac{\Gamma(-z_n)\Gamma(z_1+\cdots+z_n+\frac{d}{2})\Gamma(z_1+\cdots+z_n+\nu_n)\displaystyle\prod_{i=1}^{n-1}\Gamma(-z_i)\Gamma(\nu_i-\frac{d}{2}-z_i)}{\Gamma(z_n+\frac{d}{2})}\times\\
        &(-x_0)^{z_n}x_n^{-(z_1+\cdots+z_n+\nu_n)}\prod_{i=1}^{n-1}x_i^{\frac{d}{2}-\nu_i+z_i} dz_1\cdots dz_n\,.
    \end{split}
    \end{equation}
\end{theorem}

\begin{proof}
    It is analogous to~\cite{Vanhove:2026fth}.
    For the readers' convenience, we include its derivation.
    Firstly we define the Mellin and inverse Mellin transform as follows:
    \begin{equation}
        M[f](\zeta):=\int_0^\infty f(t)t^{-\zeta-1}dt,\quad
        M^{-1}[F](t):=\frac{1}{2\pi\ii}\int_{\ii\mathbb{R}-\varepsilon}F(\zeta)t^{\zeta}d\zeta\,,
    \end{equation}
    where $\varepsilon>0$ is a sufficiently small number.
    Setting $t:=-x_0$, we find
    \begin{equation}\label{eq:Mellin Feynman}
        M[\tilde{I}_{\nu_1,\dots,\nu_n}](\zeta)=\Gamma(-\zeta)\int_{\mathbb{R}_+^n}  \left(\sum_{i=1}^n a_i^{-1}\right)^{-\frac{d}{2}-\zeta}\exp\left\{-\sum_{i=1}^nx_i a_i\right\} a^{\nu-\frac{d}{2}{\bf 1}}\frac{d a}{a}\,.      
    \end{equation}
    Now, we use the following formula:
    \begin{equation}\label{eq:power function Mellin}
        \begin{split}
&(2\pi\ii)^{n-1}\Gamma(\zeta+\frac{d}{2})\left(\sum_{i=1}^n a_i^{-1}\right)^{-\frac{d}{2}-\zeta}\\
=&\int_{(\ii\mathbb{R}-\varepsilon)^{n-1}}\Gamma(\zeta+\frac{d}{2}+z_1+\cdots+z_{n-1}) a_n^{\zeta+\frac{d}{2}}\prod_{i=1}^{n-1}\Gamma(-z_i)\left(\frac{a_n}{a_i}\right)^{z_i}dz_i\,.
        \end{split}
    \end{equation}
    The formula~\eqref{eq:Mellin-Barnes representation} follows by substituting~\eqref{eq:power function Mellin} to~\eqref{eq:Mellin Feynman}, performing inverse Mellin transform with respect to $z_n:=\zeta$ and integrating out the $a$ variables.
\end{proof}

As in~\cite{Vanhove:2026fth}, we first shift the integration contour in $z_n$-space to the left to take residues at
$$
z_n=-z_1-\cdots-z_{n-1}-\nu_n-m_n\ \ \ (m_n=0,1,2,\dots)
$$
and
$$
z_n=-z_1-\cdots-z_{n-1}-\frac{d}{2}-m_n\ \ \ (m_n=0,1,2,\dots)\,.
$$

Then, the integral $\tilde{I}_{\nu_1,\dots,\nu_n}$ is expanded as
\begin{equation}\label{eq:Mellin Barnes expanded}
        \begin{split}
        \tilde{I}_{\nu_1,\dots,\nu_n}=&\frac{1}{(2\pi\ii)^{n-1}}\sum_{m_n=0}^\infty\int_{(\ii\mathbb{R}-\varepsilon)^{n-1}}\\
        &\frac{\Gamma(z_1+\cdots+z_{n-1}+\nu_n+m_n)\Gamma(\frac{d}{2}-\nu_n-m_n)\displaystyle\prod_{i=1}^{n-1}\Gamma(-z_i)\Gamma(\nu_i-\frac{d}{2}-z_i)}{\Gamma(-z_1-\cdots-z_{n-1}-\nu_n+\frac{d}{2}-m_n)m_n!}\times\\
        &(-x_0)^{-z_1-\cdots-z_{n-1}-\nu_n-m_n}x_n^{m_n}\prod_{i=1}^{n-1}x_i^{\frac{d}{2}-\nu_i+z_i} (-1)^{m_n}dz_1\cdots dz_{n-1}+\\
        &\frac{1}{(2\pi\ii)^{n-1}}\sum_{m_n=0}^\infty\int_{(\ii\mathbb{R}-\varepsilon)^{n-1}}\\
        &\frac{\Gamma(z_1+\cdots+z_{n-1}+\frac{d}{2}+m_n)\Gamma(-\frac{d}{2}+\nu_n-m_n)\displaystyle\prod_{i=1}^{n-1}\Gamma(-z_i)\Gamma(\nu_i-\frac{d}{2}-z_i)}{\Gamma(-z_1-\cdots-z_{n-1}-m_n)m_n!}\times\\
        &(-x_0)^{-z_1-\cdots-z_{n-1}-\frac{d}{2}-m_n}x_n^{\frac{d}{2}-\nu_n+m_n}\prod_{i=1}^{n-1}x_i^{\frac{d}{2}-\nu_i+z_i} (-1)^{m_n}dz_1\cdots dz_{n-1}\,.
        \end{split}
    \end{equation}
Now, we take residues of the integrand of each summand.
To do this, we shift the integration contour to the right in the $z_i$-plane.
The residues are at
$$
z_i=m_i\quad\text{and}\quad z_i=\nu_i-\frac{d}{2}+m_i\quad (m_i=0,1,2,\dots)\,.
$$
Here, we note that the second integrand does not produce residues when
$z_1,\dots,z_{n-1}\in\{0,1,2,\dots\}$ because of its denominator.

We pick a subset $J\subset\{ 1,2,\dots,n-1\}$ and take residues of the first integrand of~\eqref{eq:Mellin Barnes expanded} at 
$$
z_i=m_i\ \ (i\in J) \quad z_i=\nu_i-\frac{d}{2}+m_i\ \ (i\notin J)\,.
$$
It gives rise to the solution $F_J(x;\nu)$ defined by~\eqref{eq:Lauricella banana}.
Likewise, we pick a subset $J'\subset\{ 1,2,\dots,n-1\}$ and take residues of the second integrand of~\eqref{eq:Mellin Barnes expanded} at 
$$
z_i=m_i\ \ (i\in J') \quad z_i=\nu_i-\frac{d}{2}+m_i\ \ (i\notin J')\,.
$$
It gives rise to the solution $F_J(x;\nu)$ with $J:=J'\cup\{ n\}$ defined by~\eqref{eq:Lauricella banana}.
The result is the following.

\begin{theorem}\label{thm:expansion}
    In~\eqref{eq:Lauricella banana}, choose ${\rm arg}(x_0)=-\pi$, ${\rm arg}(x_i/x_0)=\pi$ ($i=1,\dots,n$) and $x_0^a=e^{-a\pi\sqrt{-1}}|x_0|^a$ for any $a\in\CC$.
    When $|x_i|<\!\!<|x_0|$ for any $i=1,\dots,n$, one has the following expansion:
    \begin{equation}\label{eq:expansion}
        \tilde{I}_{\nu_1,\dots,\nu_n} =\sum_{J\subsetneq\{ 1,\dots,n\}}C_J(\nu)F_J(x;\nu)\,,
    \end{equation}
    where $C_J(\nu)$ is a meromorphic function defined by
    \begin{equation}\label{eq:C nu}
    \begin{split}
        C_J(\nu):=&e^{\pi\sqrt{-1}\left(\frac{(n-|J|-1)d}{2}-\sum_{i\notin J}\nu_i\right)}
        \Gamma\left(\sum_{i\notin J}\nu_i-\frac{(n-|J|-1)d}{2}\right)\frac{\prod_{i\in J}\Gamma(\nu_i-\frac{d}{2})\prod_{i\notin J}\Gamma(\frac{d}{2}-\nu_i)}{\Gamma\left(\frac{(n-|J|)d}{2}-\sum_{i\notin J}\nu_i\right)}\,.
    \end{split}
    \end{equation}
\end{theorem}
The formula~\eqref{eq:expansion} is still valid for other arguments of $x_0,\dots,x_n$ after analytic continuation.

\begin{example}
    For $L=1$ (i.e., $n=2$) and $\nu_1=\nu_2=1$, we find from~\eqref{eq:Lauricella banana} that
    \begin{equation}
        \begin{split}
            \tilde{I}_{1,1} =&C_\varnothing x_0^{\frac{d}{2}-2}F_4\left(\substack{2-\frac{d}{2},3-d\\ 2-\frac{d}{2},2-\frac{d}{2}};\frac{x_1}{x_0},\frac{x_2}{x_0}\right)+C_{1}x_0^{-1}x_1^{\frac{d}{2}-1}F_4\left(\substack{1,2-\frac{d}{2}\\ \frac{d}{2},2-\frac{d}{2}};\frac{x_1}{x_0},\frac{x_2}{x_0}\right)\\
            &+C_{2}x_0^{-1}x_2^{\frac{d}{2}-1}F_4\left(\substack{1,2-\frac{d}{2}\\ 2-\frac{d}{2},\frac{d}{2}};\frac{x_1}{x_0},\frac{x_2}{x_0}\right)\,,
        \end{split}
    \end{equation}
    where
    \begin{equation}
        C_\varnothing=
        e^{\pi\sqrt{-1}(\frac{d}{2}-2)}\Gamma(2-\frac{d}{2})\frac{\Gamma(\frac{d}{2}-1)^2}{\Gamma(d-2)}
        =2^{3-d}e^{\pi\sqrt{-1}\frac{d}{2}}\sqrt{\pi}\cot\left(\frac{d\pi}{2}\right)\Gamma\left(\frac{3}{2}-\frac{d}{2}\right)\,,
        \end{equation}
        and
        \begin{equation}
        C_1=C_2=\frac{-\pi}{\sin\pi(1-\frac{d}{2})\Gamma(\frac{d}{2})}=-\Gamma(1-\frac{d}{2})\,.
        \end{equation}
\end{example}

\section{Irreducibility of the monodromy representation} \label{sec:monodromy}
As proved in~\cite[Theorem 5.2 and Theorem 6.6]{matsubara2026principal}, the singular locus of $M(\alpha)$ is contained in the vanishing locus of a polynomial $x_0\cdots x_nR_{\rm ban}(x)$ where 
\begin{equation}\label{eq:R}
R_{\rm ban}(x)=\prod_{(\epsilon_0,\dots,\epsilon_n)\in\{\pm1\}^{n+1}/\{(-1,\dots,-1),(+1,\dots,+1)\}}(\epsilon_0 \sqrt{x_0}+\cdots +\epsilon_n \sqrt{x_n})\,.    
\end{equation}
Such a locus is denoted by $V(x_0\cdots x_nR_{\rm ban})\subset \CC^{n+1}$.
Given a point $x^*\in \CC^{n+1}\setminus V(x_0\cdots x_nR_{\rm ban})$, the analytic continuation of holomorphic solutions to $M(\alpha)$ induces the monodromy representation of the fundamental group $\pi_1(\CC^{n+1}\setminus V(x_0\cdots x_nR_{\rm ban}),x^*)$, which we simply call the monodromy representation of $M(\alpha)$.
In this section, we prove the following theorem:
\begin{theorem}\label{thm:irreducible}
  If the condition 
  \begin{align}\label{eq:irred-condition-alpha}
    \sum_{j\in J} \alpha_j \notin \ZZ, \quad \emptyset \neq J \subset \brc{0,1,\dots,n}
  \end{align}
  holds, the monodromy representation of $M(\alpha)$ is irreducible.
\end{theorem}
Note that the condition~\eqref{eq:irred-condition-alpha} excludes integer propagator powers: in fact, it implies a relation $\ZZ\not\ni\alpha_0+\alpha_i=\nu_i+1$.

To prove this theorem, we use explicit expressions for the monodromy matrices. 
We utilize results for the monodromy of $F_C$ as follows. 
We fix the point $x^*=(1,\frac{1}{2n^2},\dots ,\frac{1}{2n^2}) \in \CC^{n+1}$. 
For $i=1,\dots ,n$, we set $z_i =x_i/x_0$ and let $z^*=(\frac{1}{2n^2},\dots ,\frac{1}{2n^2})$ be the point corresponding to $x^*$. 
Since $x_0\cdots x_nR_{\rm ban}(x)$ is homogeneous, the map 
\begin{align*}
  \CC^* \times (\CC^n \setminus V(z_1\cdots z_n R_{\rm ban}')) \to \CC^{n+1}\setminus V(x_0\cdots x_nR_{\rm ban}) ,\quad 
  (t,z) \mapsto t(1,z) 
\end{align*}
is a homeomorphism, where $R_{\rm ban}'(z_1,\dots ,z_n)=R_{\rm ban}(1,z_1,\dots ,z_n)$.
Hence we have an isomorphism 
\begin{align*}
  \pi_1(\CC^{n+1}\setminus V(x_0\cdots x_nR_{\rm ban}),x^*) \simeq \pi_1(\CC^{n}\setminus V(z_1\cdots z_n R_{\rm ban}'),z^*) \times \ZZ\,.
\end{align*} 
Here, a generator of the factor $\ZZ$ is given by the loop $\rho_{\infty}$ defined by $\rho_{\infty}(\theta)=e^{2\pi \ii \theta}x^*$, $\theta \in [0,1]$. 
By~\cite{Goto-FC-monodromy}, $\pi_1(\CC^{n}\setminus V(z_1\cdots z_n R_{\rm ban}'),z^*)$ is generated by 
$n+1$ loops $\rho_0,\dots,\rho_n$ such that 
\begin{itemize}
\item $\rho_0$ goes once around the hypersurface $V(R_{\rm ban}')$ in the positive direction, near the point  
  $\left( \frac{1}{n^2},\ldots, \frac{1}{n^2} \right)$, 
\item $\rho_i \ (i=1,\dots , n)$ goes once around the hyperplane $V(z_i)$ in the positive direction. 
\end{itemize}
For the explicit definitions of these loops, see~\cite{Goto-FC-monodromy}. 
We regard $\rho_0,\dots,\rho_n$ as loops in $\CC^{n+1}\setminus V(x_0\cdots x_nR_{\rm ban})$ via $z\mapsto (1,z)$. 
Thus $\pi_1(\CC^{n+1}\setminus V(x_0\cdots x_nR_{\rm ban}),x^*)$ is generated by $\rho_0,\dots,\rho_n$ and $\rho_{\infty}$. 
In the following subsections, we derive explicit expressions for the monodromy matrices $N_0,\dots ,N_n$ 
with respect to a basis obtained from~\eqref{eq:Lauricella} by multiplying by some $\Gamma$-factors, 
where $N_j$ corresponds to the analytic continuation along $\rho_j$. 
\begin{lemma}
  The monodromy group of $M(\alpha)$ is generated by $N_0,\dots ,N_n$ and the scalar matrix $e^{2 \pi \ii \beta}E_{2^n-1}$. 
  In particular, the monodromy representation of $M(\alpha)$ is irreducible if and only if the action of the group $\vev{N_0,\dots ,N_n}$ on $\CC^{2^n-1}$ is irreducible.
\end{lemma}
\begin{proof}
  Since $\pi_1(\CC^{n+1}\setminus V(x_0\cdots x_nR_{\rm ban}),x^*)$ is generated by $\rho_0,\dots,\rho_n$ and $\rho_{\infty}$, 
  the monodromy group is generated by $N_0,\dots,N_n$ and $N_{\infty}$, 
  where $N_\infty$ is the matrix representing the analytic continuation along $\rho_{\infty}$. 
  Hence it suffices to show that $N_{\infty} =e^{2 \pi \ii \beta}E_{2^n-1}$. 
  By the expression~\eqref{eq:Lauricella}, 
  every basis function is of the form $x_0^{\beta} g(z)$ with $z_i =x_i/x_0$. 
  Since $g(z)$ does not change under the analytic continuation along $\rho_{\infty}$, we have $N_{\infty}=e^{2 \pi \ii \beta} E_{2^n-1}$. 
\end{proof}
Since the solution space of $M(\alpha)$ can be regarded as a non-trivial invariant subspace of 
the reducible monodromy of Lauricella's $F_C$, we use results for the monodromy of $F_C$. 
We first review the monodromy of $F_C$ with generic parameters, for which the system is irreducible. 
Next, by taking a suitable limit of the parameters, we derive expressions for $N_0,\dots ,N_n$ 
and prove the irreducibility of the monodromy of $M(\alpha)$.

\subsection{Irreducibility of Lauricella's \texorpdfstring{$F_C$}{F\_C}}
We review the monodromy representation of Lauricella's $F_C\left(\substack{a,b \\ c_1,\dots,c_n};z\right)$. 
For a subset $J\subset \brc{1,\dots ,n}$, we set 
\begin{align}
  a^J \defas a+\sum_{j\in J}(1-c_j) ,\ 
  b^J \defas b+\sum_{j\in J}(1-c_j) ,\ 
  c_j^J \defas
  \begin{cases}
    c_j & (j \notin J) \\
    2-c_j & (j \in J).
  \end{cases}\,.
\end{align}
By~\cite{hattori-takayama-2014}, the monodromy is irreducible if and only if the following condition holds true: 
\begin{align}
  a^J, b^J \notin \ZZ ,\quad  J \subset \brc{1,\dots ,n}\,.\label{eq:7.3}
\end{align}
In this subsection, we assume the condition~\eqref{eq:7.3} and $c_j \notin \ZZ$. We also set 
\begin{align}
  \lambda_a \defas e^{2\pi \sqrt{-1} a},\ 
  \lambda_b \defas e^{2\pi \sqrt{-1} b},\ 
  \lambda_j \defas e^{2\pi \sqrt{-1} c_j} \ (j=1,\dots ,n)\,. 
\end{align}

We take the following basis of the solution space: 
\begin{align}\label{eq:FC-series-solution-Gamma}
  \frac{\prod_{j\in J} \Gamma (c_j -1) \prod_{j\notin J} \Gamma (1-c_j) \Gamma (1-a^{\brc{1,\dots,n}}) \Gamma (1-b)}{\Gamma (1-a^J) \Gamma (1-b^J)} 
  \cdot \prod_{j\in J}z_j^{1-c_j} \cdot F_C \left(\substack{a^J,b^J \\ c^J} ;z \right)\,,
\end{align}
for $J\subset \brc{1,\dots ,n}$. 
By~\cite{Goto-FC-monodromy}, 
we obtain the matrix representation of the monodromy with respect to this basis. 
The monodromy group is generated by the $2^n \times 2^n$ matrices $M_0,\dots ,M_n$ defined as follows.
Each $2^n \times 2^n$ matrix is indexed by the subsets $J$ of $\brc{1,\dots ,n}$. 
\begin{enumerate}[(i)]
\item For $j=1,\dots ,n$, the matrix $M_j$ is a diagonal matrix whose $(J,J)$-th entry is 
  $\lambda_j^{-1}$ (resp. $1$) if $J\ni j$ (resp. $J\not\ni j$). 
\item The matrix $M_0$ is defined by 
  $M_0 =E_{2^n}-\frac{(\lambda_b-1)(\lambda_a -\prod_j \lambda_j)}{\lambda_a \lambda_b}
  \cdot \mathbf{1}\cdot \mathbf{1}^T \cdot H_h$, 
  where $\mathbf{1}$ is the column vector of length $2^n$ with all entries $1$ 
  and $H_h$ is the diagonal matrix whose $(J,J)$-th entry is 
  \begin{align}\label{eq:FC-homology-intersection}
  H_{JJ}=
    (-1)^{|J|} \frac{\prod_{j\notin J}\lambda_j \cdot (\lambda_a-\prod_{j\in J}\lambda_j)(\lambda_b-\prod_{j\in J}\lambda_j)}
    {\prod_{j=1}^n (\lambda_j-1) \cdot (\lambda_a-\prod_{j=1}^n \lambda_j)(\lambda_b-1)}\,.
  \end{align}
\end{enumerate}
Note that the matrices $M_1,\dots ,M_n$ commute with each other. 
The eigenvalues of $M_0$ are $1$ and $\delta \defas (-1)^{n-1} \lambda_1 \cdots \lambda_n \lambda_a^{-1}\lambda_b^{-1}$. 
Further, $\mathbf{1}$ is a $\delta$-eigenvector of $M_0$, and 
the $1$-eigenspace $W_1$ of $M_0$ is expressed as $W_1 =\brc{ w\in \CC^{2^n} \mid w^T H_h \mathbf{1}=0}$; 
in particular, it is $(2^n -1)$-dimensional.  

In the next subsection, we consider the case when $a^{\brc{1,\dots ,n}} =1 \in \ZZ$. 
In such a case, the basis~\eqref{eq:FC-series-solution-Gamma} is not well-defined. 
Similarly to~\cite[Theorem 5.1]{GotoMatsumoto-FC}, we rescale it by $H_h^{-1}$. 
With respect to the rescaled basis, the monodromy matrices are given by $M_j' \defas H_h M_j H_h^{-1}$ for $j=0,\dots ,n$. 
It is obvious that $M_j' =M_j$ for $j=1,\dots,n$, and 
$M_0' =E_{2^n}-\frac{\lambda_b-1}{\lambda_a \lambda_b} \cdot ((\lambda_a -\prod_j \lambda_j)H_h)
\cdot \mathbf{1}\cdot \mathbf{1}^T$. 
Note that we can take the limit $a^{\brc{1,\dots ,n}} \to 1$ (equivalently, $\lambda_a -\prod_j \lambda_j \to 0$) 
of $(\lambda_a -\prod_j \lambda_j)H_h$ and its $(\brc{1,\dots,n},\brc{1,\dots,n})$-entry becomes $0$. 

\subsection{Irreducibility of \texorpdfstring{$M(\alpha)$}{M(alpha)}}
We investigate the monodromy representation of $M(\alpha)$ under the condition~\eqref{eq:irred-condition-alpha}. 
By setting $j=0$ in Theorem \ref{thm:Jn-FC}, 
this monodromy is related to that of $F_C\left(\substack{a,b \\ c_1,\dots,c_n} ;z\right)$ with\footnote{Note that $a$ and $b$ are interchanged compared with~\eqref{eq:Lauricella}, which is allowed by the symmetry of $F_C$.} 
\begin{align*}
  a=-\beta-\alpha_0 +1,\quad b=-\beta ,\quad c_j=\alpha_j \quad (j=1,\dots ,n)\,. 
\end{align*}
Under the substitution~\eqref{eq:propagators and alpha}, these parameters are also written as 
\begin{equation}\label{eq:parameter substitution}
\begin{split}
  a&= -\frac{nd}{2}+\nu_1 +\dots +\nu_n +1,\quad 
  b= -\frac{(n-1)d}{2}+\nu_1 +\cdots +\nu_n ,\\
  c_j &= \nu_j +1-\frac{d}{2} \quad (j=1,\dots ,n)\,.    
\end{split}
\end{equation}
Since $a^{\brc{1,\dots ,n}} =1 \in \ZZ$ by the constraint~\eqref{eq:constraint}, the monodromy representation of $F_C$ is reducible. 
Note that we have $a^J,b^J \notin \ZZ$ ($J\subsetneq \brc{1,\dots ,n}$), $c_j\notin \ZZ$ ($j=1,\dots ,n$) and $b^{\brc{1,\dots ,n}}\not\in \ZZ$.
We use the matrix representation $M_0',\dots,M_n'$ of the monodromy. 
By taking the limit $a^{\brc{1,\dots ,n}} \to 1$, we have 
\begin{align*}
  &M_j' (=M_j)\to \lrbrk{
  \begin{array}{c|c}
    N_j & * \\ \hline
    0 \cdots 0 & \lambda_j^{-1}
  \end{array}
  } \quad (j=1,\dots ,n) ,\\
  &M_0' \to \lrbrk{
  \begin{array}{c|c}
    N_0 & * \\ \hline
    0 \cdots 0 & 1
  \end{array}
  }\,,
\end{align*}
where $N_0,\dots ,N_n$ are $(2^n -1)\times (2^n -1)$ matrices. 
This implies that 
$\brc{ v\in \CC^{2^n} \mid v_{\brc{1,\dots, n}} =0 } \simeq \CC^{2^n-1}$ is 
a non-trivial invariant subspace, and hence we verify that the monodromy representation is reducible.  

\begin{proposition}
  For $J\subsetneq \brc{1,\dots ,n}$, 
  we set
  \begin{align*}
  B_J(\alpha)&:=\frac{\Gamma(|\alpha_{\bar{J}}|-(n-|J|-1))\Gamma(\alpha_0+|\alpha_{\bar{J}}|-(n-|J|))}{\prod_{j\in J}\Gamma(2-\alpha_j)\prod_{j\notin J}\Gamma(\alpha_j)}\\
  &=\frac{\Gamma(-\frac{(n-|J|)d}{2}+|\nu_{\bar{J}}|+1)\Gamma(-\frac{(n-|J|-1)d}{2}+|\nu_{\bar{J}}|)}{\prod_{j\in J}\Gamma(1+\frac{d}{2}-\nu_j)\prod_{j\notin J}\Gamma(1-\frac{d}{2}+\nu_j)}\,,
  \end{align*}
  which is well-defined and nonzero under the assumption~\eqref{eq:irred-condition-alpha}. 
  Let $F_J$ be the solution to $M(\alpha)$ given by~\eqref{eq:Lauricella} for $j=0$ 
  (or equivalently,~\eqref{eq:Lauricella banana}). 
  Then $N_0 ,\dots ,N_n$ are 
  the monodromy matrices of $M(\alpha)$ with respect to the basis 
  $\brc{ B_J(\alpha)F_J }_{J\subsetneq \brc{1,\dots ,n}}$. 
\end{proposition}
\begin{proof}
    By construction, $N_0 ,\dots ,N_n$ are the monodromy matrices of $M(\alpha)$ with respect to the basis 
    \begin{align*}
      \left. 
      \frac{1}{H_{JJ}}\cdot 
      \frac{\prod_{j\in J} \Gamma (c_j -1) \prod_{j\notin J} \Gamma (1-c_j) \Gamma (1-a^{\brc{1,\dots,n}}) \Gamma (1-b)}{\Gamma (1-a^J) \Gamma (1-b^J)} 
      \right|_{\begin{subarray}{l} a=-\beta-\alpha_0 +1 \\b=-\beta \\c_j=\alpha_j \end{subarray}} 
      \cdot F_J\,.
    \end{align*}
A straightforward calculation shows that this basis equals $((-2\pi\ii)^n/\Gamma(-\beta))\cdot B_J(\alpha)F_J$.
Since this factor is nonzero and independent of $J$, the claim follows.
\end{proof}

To prove Theorem \ref{thm:irreducible}, it suffices to show that 
the action of the group $\vev{N_0 ,\dots ,N_n}$ on $\CC^{2^n-1}$ is irreducible. 
We derive explicit expressions for $N_0 ,\dots ,N_n$. 
Each $(2^n -1)\times (2^n -1)$ matrix is indexed by $J\subsetneq \brc{1,\dots ,n}$. 
It is easy to see that $N_j$ ($j=1,\dots ,n$) is a diagonal matrix whose $(J,J)$-th entry is 
$\lambda_j^{-1}$ (resp. $1$) if $J\ni j$ (resp. $J\not\ni j$). 
Let $H_{h0}$ be the $(2^n -1)\times (2^n -1)$ matrix defined by 
\begin{align*}
  \lim_{a^{\brc{1,\dots ,n}} \to 1} \Bigbrk{ (\lambda_a -\prod_j \lambda_j)H_h }
  =\lrbrk{
  \begin{array}{c|c}
    H_{h0} &  \\ \hline
     & 0
  \end{array}
  }\,.
\end{align*}
Then $N_0$ is expressed as 
\begin{equation}\label{eq:N_0}
    N_0=E_{2^n-1}-\frac{\lambda_b-1}{\lambda_a \lambda_b} \cdot H_{h0} \cdot \mathbf{1}_0\cdot \mathbf{1}_0^T\,,
\end{equation} 
where $\mathbf{1}_0$ is the column vector of length $2^n-1$ with all entries $1$. 
Since $\bigbrk{ (\lambda_a -\prod_j \lambda_j)H_h }\mathbf{1}$ is an eigenvector of $M_0'$ with eigenvalue $\delta$, 
we can check that $h\defas H_{h0}\mathbf{1}_0$ is an eigenvector of $N_0$ with eigenvalue 
$\delta (=(-1)^{n-1}\lambda_b^{-1}=(-1)^{n-1}e^{2\pi \sqrt{-1}\beta})$. 
Moreover, the $1$-eigenspace $V_1$ of $N_0$ is characterized as 
\begin{align}\label{eq:eigenspace-V1}
  V_1 =\brc{ v\in \CC^{2^n-1} \mid v^T H_{h0}^{-1} h =0}
  =\brc{ v\in \CC^{2^n-1} \mid v^T \mathbf{1}_0 =0}\,.
\end{align}

For $J\subset \brc{1,\dots ,n}$, we set $N_J =\prod_{j\in J}N_j$. 
Here, we note that $N_1,\dots ,N_n$ commute with each other. 
\begin{lemma}\label{lemma:NJ1-lin-indep}
  $\brc{N_J \mathbf{1}_0 }_{J\subsetneq \brc{1,\dots ,n}}$ is linearly independent.
\end{lemma}
\begin{proof}
  Let $A$ be the $(2^n -1)\times (2^n -1)$ matrix defined by $A=(\cdots , N_J \mathbf{1}_0 ,\cdots)$, where the index $J$ runs over proper subsets $J\subsetneq \brc{1,\dots ,n}$. It suffices to show that $\det A\ne 0$.
  We consider the $2^n \times 2^n$ matrix $\tilde{A}$
  whose $(J_1,J_2)$-th entry is $\prod_{j\in J_1 \cap J_2} \lambda_j^{-1}$, 
  where the indices $J_1,J_2$ run over the subsets of $\brc{1,\dots ,n}$. 
  Since $\tilde{A}$ is written as
  \begin{align*}
    \tilde{A}=
    \lrbrk{ \begin{array}{cc} 1 & 1 \\ 1 & \lambda_1^{-1} \end{array} }
    \otimes \dots \otimes
    \lrbrk{ \begin{array}{cc} 1 & 1 \\ 1 & \lambda_n^{-1} \end{array} }\,,
  \end{align*}
  we have $\det \tilde{A}=\prod_{j=1}^n (\lambda_j^{-1} -1)^{2^{n-1}}$. 
  It is easy to see that the submatrix of $\tilde{A}$ obtained by deleting the $\brc{1,\dots ,n}$-th row and column is equal to $A$.
  We apply the following elementary row operations to $\tilde{A}$:
  first, multiply the $\brc{1,\dots ,n}$-th row by $(-1)^n$, and then,
  for each $J\subsetneq \brc{1,\dots ,n}$, 
  add $(-1)^{|J|}$ times the $J$-th row to the $\brc{1,\dots ,n}$-th row.
  Then the determinant of $\tilde{A}$ is also computed as 
  \begin{align*}
    \det \tilde{A}
    =(-1)^n \cdot \det
    \lrbrk{
    \begin{array}{c|c}
      A & * \\ \hline
      0\cdots 0 & \prod_{j=1}^n (1-\lambda_j^{-1})
    \end{array}
    }
    =\det A \cdot \prod_{j=1}^n (\lambda_j^{-1}-1)\,. 
  \end{align*}
  Therefore, it follows that $\det A=\prod_{j=1}^n (\lambda_j^{-1} -1)^{2^{n-1}-1}$. 
  Since each $c_j$ ($j=1,\dots ,n$) is not an integer, we obtain $\det A\ne 0$. 
\end{proof}

\begin{corollary}\label{cor:NJh-lin-indep}
  $\brc{N_J h}_{J\subsetneq \brc{1,\dots ,n}}$ is linearly independent.  
\end{corollary}
\begin{proof}
  Since $N_J$ commutes with $H_{h0}$, we have 
  $(\cdots , N_J h,\cdots)=H_{h0}(\cdots , N_J \mathbf{1}_0 ,\cdots)$. 
  Then the determinant of $(\cdots , N_J h,\cdots)$ is not zero. 
\end{proof}

Theorem \ref{thm:irreducible} is a consequence of the following theorem.
\begin{theorem} 
  Under the assumptions $c_j \notin \ZZ$ and~\eqref{eq:7.3} but with $a^{\{1,\dots,n\}}\in \mathbb Z$ instead, the action of the group $\vev{N_0,\dots ,N_n}$ on $\CC^{2^n-1}$ is irreducible. 
\end{theorem}
\begin{proof}
  Let $V \subset \CC^{2^n-1}$ be an invariant subspace. 
  \begin{enumerate}[(i)]
  \item Suppose that $V \not\subset V_1$. 
    By the expression~\eqref{eq:eigenspace-V1}, there exists $v\in V$ such that $v^T \mathbf{1}_0 \neq 0$. 
    Since 
    \begin{align*}
      N_0 v =v-\frac{\lambda_b-1}{\lambda_a \lambda_b} \cdot H_{h0} \cdot \mathbf{1}_0\cdot \mathbf{1}_0^T \cdot v
      =v-\frac{(\lambda_b-1)\cdot (v^T \mathbf{1}_0)}{\lambda_a \lambda_b} \cdot h
    \end{align*}
    belongs to $V$, we can show that $h$ also belongs to $V$. 
    Thus we have $N_J h \in V$ for each $J\subsetneq \brc{1,\dots ,n}$. 
    Corollary \ref{cor:NJh-lin-indep} implies that $\dim V \geq 2^n -1$, and hence we obtain $V=\CC^{2^n-1}$. 
  \item Suppose that $V \subset V_1$. 
    We fix an arbitrary $v \in V$. 
    For any $J\subset \brc{1,\dots ,n}$, we have $N_J v \in V$, which implies
    $v^T \cdot (N_J \mathbf{1}_0)=(N_J v)^T \cdot \mathbf{1}_0 =0$. 
    By Lemma \ref{lemma:NJ1-lin-indep}, $\brc{N_J \mathbf{1}_0 }_{J\subsetneq \brc{1,\dots ,n}}$ is 
    a basis of $\CC^{2^n-1}$, and hence we obtain $v=0$. 
    Therefore, we have $V=\brc{0}$.
  \end{enumerate}
  By (i) and (ii), every invariant subspace is trivial. 
\end{proof}

\subsection{Discontinuity integrals}
Finally, we relate the analysis above with a formula for the discontinuity integral.
The most interesting component of the singular locus of $M(\alpha)$ is $V(R_{\rm ban})$.
We pick a smooth point $x^*\in V(R_{\rm ban})$ and fix a simple loop $\rho$ which goes around the component $V(R_{\rm ban})$ in a neighborhood of $x^*$.
Let $\widetilde{\mathcal{O}}_{x^*}$ be the space of multivalued holomorphic functions on a neighborhood of $x^*$ ramified only along $V(R_{\rm ban})$ and let $\mathcal{M}_{x^*}$ be the space of meromorphic functions on a neighborhood of $x^*$ whose poles are only along $V(R_{\rm ban})$.
For any holomorphic solution $f$ to $M(\alpha)$, an analytic continuation of a function $\rho_*f-f$ to a neighborhood of $x^*$ defines a well-defined element of the quotient $\widetilde{\mathcal{O}}_{x^*}/\mathcal{M}_{x^*}$.
The space of discontinuity integrals is a subspace of the quotient $\widetilde{\mathcal{O}}_{x^*}/\mathcal{M}_{x^*}$ spanned by $\rho_*f-f$ where $f$ runs over the holomorphic solutions to $M(\alpha)$.
The space of discontinuity integrals depends on the choice of a simple loop $\rho$.
However, any pair of simple loops give rise to isomorphic spaces of discontinuity integrals as any other simple loop is conjugate to $\rho$ in the fundamental group~\cite[Theorem 13.1]{HaraokaBook}.

\begin{theorem}\label{th: ban_combination}
    Under the assumption~\eqref{eq:irred-condition-alpha} and $\delta\neq 1$, the space of discontinuity integrals is a one-dimensional vector space spanned by the equivalence class of the sum
    \begin{equation}
        \sum_{J\subsetneq\{1,\dots,n\}}\tilde{C}_J(\nu)F_J(x;\nu)\,,
    \end{equation}
    where 
\begin{equation}\label{eq: tilde C}
\tilde{C}_J(\nu)=\frac{\prod_{j\notin J}\Gamma(\frac{d}{2}-\nu_j)\prod_{j\in J}\Gamma(\nu_j-\frac{d}{2})}{\Gamma(\frac{(n-|J|)d}{2}-|\nu_{\bar{J}}|)\Gamma(1+\frac{(n-|J|-1)d}{2}-|\nu_{\bar{J}}|)}\,.
\end{equation}
\end{theorem}
\begin{proof}
The space of discontinuity integrals is identified with the image of the operator $N_0-1$ in the quotient space $\widetilde{\mathcal{O}}_{x^*}/\mathcal{M}_{x^*}$, which is at most one-dimensional.
We choose the simple loop to be $\rho_0$.
If the space of discontinuity integrals is zero, this means that the monodromy along $\rho_0$ is trivial by the assumption~\eqref{eq:irred-condition-alpha} and $\delta\neq 1$.
Since the matrices $N_1,\dots,N_n$ are simultaneously diagonalizable, the monodromy representation is reducible, which contradicts Theorem \ref{thm:irreducible}.
Hence, the space of discontinuity integrals is one-dimensional.
By~\eqref{eq:N_0}, it is spanned by the vector $h=H_{h0}{\bf 1}_0$.
The rest of the proof is a simple computation comparing~\eqref{eq:Lauricella banana},~\eqref{eq:FC-series-solution-Gamma}, and~\eqref{eq:parameter substitution}.
\end{proof}

\begin{theorem}
    \begin{equation}\label{eq:disc_rho_0}
        {\rm Disc}_{\rho_0}(\tilde{I}_{\nu_1,\dots,\nu_n})=C(\nu)\cdot\sum_{J\subsetneq\{1,\dots,n\}}\tilde{C}_J(\nu)F_J(x;\nu)\,,
    \end{equation}
    where
    \begin{equation}
        C(\nu)=-2\pi\sqrt{-1}\delta =2\pi\sqrt{-1}(-1)^{n}e^{2\pi \sqrt{-1}\beta}\,.
    \end{equation}
\end{theorem}
\begin{proof}
    Theorem \ref{thm:expansion} expands the integral $\tilde{I}_{\nu_1,\dots,\nu_n}$ into a linear combination of the functions $\{B_J(\alpha)F_J(x;\nu)\}_J$ with coefficients
    \begin{equation*}
    \left( (-1)^{n-|J|}\pi^{n-1}e^{\pi\sqrt{-1}(\frac{(n-|J|-1)d}{2}-\sum_{i\notin J}\nu_i)}\frac{\sin\pi(\frac{(n-|J|)d}{2}-\sum_{i\notin J}\nu_i)}{\prod_{i=1}^n\sin\pi(\nu_i-\frac{d}{2})}\right)_J\,.
    \end{equation*}
    In view of the matrix representation~\eqref{eq:N_0} of $N_0$, the formula~\eqref{eq:disc_rho_0} follows.
\end{proof}

\section{Pfaffian system}  \label{sec:pfaffian system}

So far, we have discussed systems of higher order differential operators that annihilate banana integrals, which form a $D$-ideal $\mathcal{I}$. However, it is most common in physics to use a system of first order differential equations written in matrix form, the so-called Pfaffian system
\begin{align}
d I = A \, I\,, 
\end{align}
where $I=(I_{1}, \hdots \, I_{r})^T$ is a vector of basis integrals called master integrals (MI, see, e.g., a textbook~\cite{Smirnov:2012gma}), 
and $A(x,d)=\sum_i A_i dx_i$ is an $r \times r$ connection matrix
satisfying $dA-A \wedge A=0$. 
We call $A_i$'s Pfaffian matrices in this paper.
The expression using the exterior derivative is equivalent to the condition that $I$ satisfies the system of differential equations $\frac{\partial I}{\partial x_i}=A_i I$ of rank $r$.
Furthermore, when there exist suitable square free polynomials $\beta_k$ such that the connection matrix can be written as $A=\sum_k {\tilde A}_k d\log \beta_k = \sum_k {\tilde A}_k \frac{d \beta_k}{\beta_k}$, this is called a logarithmic connection matrix.

To describe the relationship between the Pfaffian system discussed in the field of physics and the system of equations $M(\alpha)=D/J_n(\alpha)$ that we have been discussing, we utilize a system of differential-difference equations.

\subsection{Differential-Difference Equations}

Let us introduce the ring of differential-difference operators
\begin{equation} \label{eq:differential-difference operators}
{\tilde D}=\CC \langle x_0, x_1, \ldots, x_n, \nu_1, \ldots, \nu_m, \partial_0, \partial_1, \ldots, \partial_n, S_1, \ldots, S_m \rangle\,. 
\end{equation}
Here, we have non-commutative relations
$\pd{i} x_j = x_j \pd{i} + \delta_{ij}$ and 
$S_i \nu_j = (\nu_j + \delta_{ij}) S_i$
where $\delta_{ij}$ is Kronecker's delta.
The $x_i$ and $\nu_j$ ($ i=0, \ldots, n, j=1, \ldots, m$)
commute with each other, and the same holds for the $\pd{i}$ and $S_j$,
$\pd{i}$ and $\nu_j$, $S_i$ and $x_j$.
The action on a function $f(\nu,x)$ is defined as 
\begin{equation} \label{eq:action of DD to a function}
 x^{p} \nu^{q} \pd{}^{p'} S^{q'} \bullet f(\nu,x)
=  x^{p} \nu^{q} \frac{\partial^{p'} f(\nu+q',x)}{\partial x^{p'}}\,.
\end{equation}
We will call ${\tilde D}$ the {\it DD Weyl algebra}
\footnote{``DD'' stands for Differential-Difference. 
To define the concept of a holonomic ideal, it is necessary to consider a ring that also includes $S_i^{-1}$; however, the concept of holonomicity is not required for the following discussion.}.
When we wish to specify the number of variables, 
we denote this ring by ${\tilde D}_{n,m}$.

The rational DD Weyl algebra can be defined in a manner analogous to the definition of the rational Weyl algebra from the Weyl algebra.
See, e.g., a textbook~\cite[Chap 6]{dojo}, on the rational Weyl algebra,
which is also called the ring of differential operators with rational function coefficients.
We denote by ${\tilde R}$ the rational DD Weyl algebra
\begin{equation} \label{eq:rational DD Weyl algebra}
{\tilde R}=\CC (x_0, x_1, \ldots, x_n, \nu_1, \ldots, \nu_m) \langle \partial_0, \partial_1, \ldots, \partial_n, S_1, \ldots, S_m \rangle\,. 
\end{equation}

Let $\mathcal{I}$ be a left ideal of the DD Weyl algebra.
The holonomic rank of $\mathcal{I}$ or ${\tilde R}\mathcal{I}$ is the dimension of
${\tilde R}/{\tilde R}\mathcal{I}$ as a vector space over
the rational function field 
$\CC(x,\nu)=\CC(x_0, \ldots,x_n, \nu_1, \ldots, \nu_m)$.
We assume the holonomic rank of $\mathcal{I}$ is finite in the sequel.
Let $b_1, \ldots, b_r$ be a basis of the vector space
\begin{equation}
\frac{{\tilde R}}{{\tilde R}\mathcal{I}}\,,
\end{equation}
where $r={\rm rank}(\mathcal{I})$ is the holonomic rank.
Then, there exist rational functions $p_{ij}^k$ and $q_{ij}^k$ satisfying
\begin{eqnarray*}
\pd{i} b_j &\equiv& \sum_{k=1}^r p_{ij}^k b_k, \quad {\rm mod}\, {\tilde R}\mathcal{I}, \\
S_i b_j &\equiv& \sum_{k=1}^r q_{ij}^k b_k, \quad {\rm mod}\, {\tilde R}\mathcal{I}\,. 
\end{eqnarray*}
Fix $i$. We call the $r \times r$ matrix $(p_{ij}^k)$ a Pfaffian matrix
and $(q_{ij}^k)$ a difference Pfaffian matrix.
Pfaffian matrices can be obtained by the normal form algorithm 
for a Gr\"obner basis\footnote{See, e.g., a textbook~\cite[Chap 6]{dojo}.  The discussion of Gr\"obner bases in the rational DD Weyl algebra is essentially the same as that in the rational Weyl algebra.}
for ${\tilde R}\mathcal{I}$
or Macaulay matrix computation~\cite{Chestnov:2022alh}.

Let us consider the banana integral $I_\nu=I_{\nu_1, \ldots, \nu_n}$.
It follows from the integral in the momentum space (\ref{eq:momentum_space_integral}) that
we have 
\begin{equation}
\frac{\partial I_\nu}{\partial x_i} = \nu_i I_{\nu+e_i}, \quad i=1, \ldots, n\,. 
\end{equation} 
In other words, $(\pd{i} - \nu_i S_i) \bullet I_\nu = 0$.

Keeping this in mind, we consider the left ideal ${\tilde J}_n(\alpha)$ generated by $J_n(\alpha)$ and $S_i - \nu_i ^{-1}\pd{i}$ for $i=1, \ldots, n$ 
in the rational DD Weyl algebra ${\tilde R}_{n,n}$ ($m=n$).
Note that $\alpha$ depends on $d$ and $\nu$ as (\ref{eq:propagators and alpha}).
In physics, master integrals  are a set of integrals $I_\nu$ for which the indices $\nu$ are chosen to form a basis for a Pfaffian system.
In other words, using the terminology of the DD Weyl algebra introduced here, if the basis elements $b_i$ are chosen as monomials in $S_j$ corresponding to the master integrals, the resulting Pfaffian matrices coincide with the Pfaffian matrices in physics mentioned at the beginning.

\begin{theorem}
The holonomic rank of the left ideal ${\tilde J}_n(\alpha)$ is $2^n-1$.
\end{theorem}

\begin{proof}
We compute the Gr\"obner basis of ${\tilde J}_n(\alpha)$ in the rational DD Weyl algebra ${\tilde R}_{n,n}$
with respect to an order such that $S_1, \ldots, S_n \succ \partial_0, \ldots, \partial_n$. 
Since the elements of $J_n(\alpha)$ do not contain $S_i$
and
the leading term of $S_i - \nu_i^{-1} \pd{i}$ is $S_i$,
we obtain a Gr\"obner basis of ${\tilde J}_n(\alpha)$ 
with at most $2^{n}-1$ standard monomials 
by the results of Sections \ref{sec:the_upper_bound} and   \ref{sec:the_lower_bound_of_the_holonomic_rank}.
From this, the upper bound of the holonomic rank is less than or equal to $r=2^{n}-1$. 

Let us give a lower bound of the holonomic rank.
As a preliminary step, we rewrite the solution constructed in the proof of Theorem \ref{thm:lower bound} for the case $j=0$ using the $\Gamma$ function.
By utilizing the relationship between the $\Gamma$ function and the Pochhammer symbol, $(p)_m = \Gamma(p+m)/\Gamma(p)$, it can be seen that these solutions are constant multiples of the following series:
\begin{align}
&F_J(d,\nu;x)=N(\nu)  \\
&\cdot \sum_{m \in \ZZ_{\geq 0}^n} \frac{\Gamma(A+\vert{}m\vert{}) \Gamma(A-\alpha_0+1+\vert{}m\vert{})}{\prod_{k\in J}\Gamma(2-\alpha_k+m_k)\prod_{k\notin J}\Gamma(\alpha_k+m_k)\prod \Gamma(m_k+1)} x_0^{-A-\vert{}m\vert{}}\prod_{k\in J}x_k^{m_k+1-\alpha_k}\prod_{k\notin J}x_k^{m_k}\,, \nonumber
\end{align}
where $A = -\beta + \vert{}J\vert{} - \sum_{k\in J} \alpha_k$
and $N(\nu) = \prod_{k=1}^n \frac{1}{\Gamma(\nu_k)}$.
That this solution is annihilated by $S_i - \nu_i^{-1}\partial_i$ (for $1 \leq i \leq n$) follows readily from the fact that $\beta$ and $\alpha_i$ are linear expressions  (\ref{eq:constraint}), (\ref{eq:propagators and alpha}) in $d$ and $\nu$, combined with the recurrence relation for the $\Gamma$ function, $\Gamma(z+1)=z\Gamma(z)$.

Let us assume that the rank of ${\tilde R}/{\tilde J}_n(\alpha)$ is less than $r$ and derive a contradiction.
By this assumption, for any set of monomials $\{b_1, \ldots, b_r\}$ in the rational DD Weyl algebra,
there exist non-trivial coefficients $c_j \in \CC(\nu, x)$ such that
$
\sum_j c_j b_j \equiv 0 \quad {\rm mod}\, {\tilde J}_n(\alpha).
$
Now, let $f_1, \ldots, f_r$ be the basis of solutions constructed above using the $\Gamma$ function.
For any value $d$ and a generic value vector $\nu$, these functions are linearly independent over $\CC$ and are locally holomorphic in the neighborhood of certain points in the $(\nu, x)$ space.
The following relation holds:
$$
\sum_{j=1}^r c_j b_j \bullet f_i = 0, \quad i=1, \ldots, r\,.
$$
For the monomials $b_j$, we take the basis provided in Theorem \ref{prop:rank_upper_bound}
(replace $\xi_i$ by $\pd{i}$ and then the basis becomes a basis of $R/RJ_n(\alpha)$
where $R$ is the rational Weyl algebra over $\CC(x_0, \ldots, x_n)$).
Defining the matrix ${\tilde W} = (b_j \bullet f_i)$ and the vector $C = (c_1, \ldots, c_r)^T$, we have ${\tilde W} C = 0$.
The determinant $W = \det({\tilde W})$ is the Wronskian of the linearly independent solutions; thus, it is non-zero at certain points. Furthermore, the $f_i$ are locally holomorphic functions of $\nu$ and $x$.
Consequently, $C$ must be zero as a vector of locally holomorphic functions of $\nu$ and $x$, which implies $C = 0$ as an element of $\CC(\nu, x)^r$.
It is a contradiction.
Therefore, the rank is $r$.
\end{proof}

When viewed from the perspective of the DD Weyl algebra,
contiguity relations in the theory of special functions
and IBP relations in physics are nothing other than elements of left ideals of the DD Weyl algebra.

\begin{example}\rm
Consider the $1$-loop case ($n=2$).
The holonomic rank is $3$.
When we compute the Gr\"obner basis $G$ for ${\tilde J}_2(\alpha)$ using the ordering employed in the theorem above,
we find that $(\pd{1}, \pd{2}, 1)$ can be taken as a basis for ${\tilde R}/{\tilde J}_2(\alpha)$.
Furthermore, if we calculate the Gr\"obner basis using the ordering $\pd{0}, \pd{1}, \pd{2} \succ S_1, S_2$,
we see that $(S_1S_2, S_1, S_2)$ also serves as a basis.

Calculating the normal forms of the elements of $(S_1S_2, S_1, S_2)$ and $\pd{0}(S_1S_2, S_1, S_2)$
with respect to the Gr\"obner basis $G$ yields
matrices $G_1$ and $G_2$ such that
\begin{eqnarray*}
(S_1S_2,S_1,S_2)^T &\equiv& G_1 (\pd{1},\pd{2},1)^T
\ {\rm mod}\, {\tilde J}_n(\alpha), \\
\pd{0}(S_1S_2,S_1,S_2)^T &\equiv& G_2 (\pd{1},\pd{2},1)^T
\ {\rm mod}\, {\tilde J}_n(\alpha)\,.
\end{eqnarray*}
For instance, regarding $G_1$,
since we can obtain $c_0, c_1, c_2 \in \CC(x,\nu)$ by the normal form algorithm such that 
$S_1S_2 \equiv \frac{1}{\nu_1 \nu_2} \pd{1}\pd{2} \equiv c_0 + c_1 \pd{1} + c_2 \pd{2}$ ${\rm mod}\, {\tilde J}_n(\alpha)$,
we have
$$
G_1=\left(\begin{array}{ccc}
c_1 & c_2 & c_0 \\ 1/\nu_1 & 0 & 0 \\
0 & 1/\nu_2 & 0 \\
\end{array}\right)\,.
$$
Therefore,
$ \pd{0}(S_1S_2,S_1,S_2)^T \equiv G_2 G_1^{-1} (S_1S_2,S_1,S_2)^T$
holds.
In other words,
if we set
$$Z=(I_{\nu_1+1,\nu_2+1}, I_{\nu_1+1,\nu_2}, I_{\nu_1,\nu_2+1})^T\,,$$
then
$ \pd{0}\bullet Z = G_2 G_1^{-1} Z$
holds.
By explicitly calculating $G_2 G_1^{-1}$, one can see that
$\nu_i$ does not appear as a factor in the denominators of the elements of $G_2 G_1^{-1}$.

In this representation, it is possible to set $\nu_i=0$.
Then, we obtain a Pfaffian system well-known in physics
for master integrals $I_{11}, I_{10}, I_{01}$
(see, e.g.,~\cite[\S 3]{abreu2022sagex}).
It is also possible to restrict it to $x_1=x_2=1$.
In the equal mass case ($x_1=x_2=1$), this Pfaffian matrix at $i=0$ is given by
$$A_0= \left(\begin{array}{ccc}   - \frac{ 1} {   2  {x}_{0}}+ \frac{   {d}- 3} {   2  (  {x}_{0}- 4)} &    -\frac{   2- {d}} {   8  {x}_{0}}- \frac{   {d}- 2} {   8  (  {x}_{0}- 4)} &    -\frac{   2- {d}} {   8  {x}_{0}}- \frac{   {d}- 2} {   8  (  {x}_{0}- 4)} \\   0&0&0 \\   0&0&0 \\ \end{array}\right)\,.$$
That is, $\pd{0}\bullet Z=A_0 Z$.
Since $I_{10}=I_{01}$, we obtain a Pfaffian system of rank $2$. 
\end{example}

Although the calculation in this example utilized the intermediate basis $(\partial_1, \partial_2, 1)$,
one could alternatively compute the normal form of $\partial_i(S_1S_2, S_1, S_2)$ directly
using a Gr\"obner basis $G'$ corresponding to the ordering 
$\{ \partial_i \} \succ \{ S_j \}$.
The optimal method should be selected by comparing the complexity of the Gr\"obner bases $G$ and $G'$
against the computational cost of the inverse matrix calculation.
We also note that an advantage of utilizing the intermediate basis is that it allows for the derivation of a Pfaffian system with respect to an arbitrary basis, not necessarily one consisting of standard monomials.

While deriving a Pfaffian system using the method demonstrated in this example is theoretically possible
for any number of loops $n-1$, 
the computation becomes increasingly difficult as the rank rises exponentially.
Determining whether a Pfaffian system can be computed for an arbitrary number of loops
via a theoretical approach remains a topic for future study.
As discussed in the next subsection, this Pfaffian system takes on a complicated form
when all $x_i$ are distinct.

\subsection{Non-existence of a logarithmic Pfaffian system}
The following theorem is an analogue of~\cite[Proposition 4.2]{GotoKanekoMatsumoto}.

\begin{theorem}\label{thm:impossible}
    Let $R(x)\in\mathbb{C}[x_0,\dots,x_n]$ be a non-constant polynomial homogeneous of degree $d$ and let  $A_0,\dots,A_{n+1}\in M(r;\mathbb{C})$ be constant matrices.
    Assume that the Newton polytope of $R(x)$ is a scaled simplex, i.e., it is $d\Delta$ where $\Delta=\{ (t_0,\dots,t_n)\in\mathbb{R}^{n+1}\mid t_0+\cdots +t_n=1, t_0,\dots,t_n\geq 0\}$ is the $n$-dimensional simplex.
    Assume $n,r\geq 2$.
    Then, the monodromy representation of the Pfaffian system
    \begin{equation}
        dY=(A_0d\log x_0+\cdots+A_nd\log x_n+A_{n+1}d\log R(x))Y=:\omega Y
    \end{equation}
    is reducible if the integrability condition is satisfied.
    More precisely, there is an invertible matrix $P\in GL(r;\mathbb{C})$ such that $P^{-1}A_iP$ is upper triangular for any $i=0,\dots,n+1$.
\end{theorem}

\begin{proof}
    By the integrability condition, we obtain that $\omega\wedge\omega=0$.
    This is equivalent to the following condition: for any $0\leq i< j\leq n$, the identity
    \begin{equation}\label{eq:ij}
        [A_i,A_j]R(x)+[A_i,A_{n+1}]x_j\frac{\partial R}{\partial x_j}(x)-[A_j,A_{n+1}]x_i\frac{\partial R}{\partial x_i}(x)=0\,.
    \end{equation}
    In fact,~\eqref{eq:ij} is the coefficient of $dx_i\wedge dx_j$ in $\omega\wedge\omega$ multiplied by $x_ix_jR(x)$.
    Now, we fix any $0\leq i<j\leq n$ and choose any $0\leq k\leq n$ so that $k\neq i,j$.
    Then, the coefficient of $x_k^d$ in~\eqref{eq:ij} is $[A_i,A_j]$ by the assumption that the Newton polytope of $R(x)$ is a scaled simplex.
    Thus, we obtain that $[A_i,A_j]=0$ for any $0\leq i<j\leq n$.
    Then, the coefficient of $x_j^d$ in~\eqref{eq:ij} is $d[A_i,A_{n+1}]$.
    Thus, it follows that $[A_i,A_{n+1}]=0$ for any $0\leq i\leq n$.
\end{proof}

For the banana integral the singular locus is given by $V(x_0\cdots x_nR_{\rm ban}(x))$ where $R_{\rm ban}(x)$ is given by~\eqref{eq:R}.
It is readily seen from this formula that the assumption of Theorem \ref{thm:impossible} is fulfilled.
We remark that one may find a logarithmic connection matrix after a covering transformation of coordinates.
For example, this is explicitly achieved in~\cite[Theorem 1.1]{kato1995connection} for Appell's $F_4$ system, which amounts to $n=2$ in our setup.

\section{The \texorpdfstring{$d\log$}{dlog} basis for equal mass banana integrals} \label{sec:equalmass}

In this section, we discuss the Pfaffian system for the equal-mass case based on Picard-Fuchs equations for period integrals of $(n-2)$-dimensional Calabi--Yau varieties.
For a given choice of master integrals,
we prove that the system is in $d\log$ form
at every loop order.

Set all internal masses equal, $x_{1}=\dots=x_{n}=m^{2}$, and write
\begin{equation}
\label{eq:em_setup}
    x=\frac{x_0}{x_1} \,, \qquad n=L+1 \,, \qquad
    c=\Bigl\lceil \frac{n}{2}\Bigr\rceil \,.
\end{equation}
On the equal-mass locus, the banana integral family at $L$ loops has $n$ master integrals~\cite{Kalmykov:2016lxx,Bonisch:2020qmm,Pogel:2022vat}, and the singular locus is given by
\begin{equation}\label{eq:em_letters}
    \beta_{0}=x \,, \qquad
    \beta_{k}=x-s_{k} \,, \quad s_{k}=(n-2k+2)^{2} \,, \quad k=1,\dots,c \,,
\end{equation}
together with $x=\infty$: the letters are the thresholds and pseudo-thresholds
$x_0=(n-2k+2)^{2}m^{2}$, and there are $c+1$ of them.

Its annihilator is generated by a single operator $\mathcal{L}_{n}$ of order $n$, that factorises as
\begin{equation}\label{eq:em_LdK}
    \mathcal{L}_{n}=\partial_{x}\circ K_{L} \,,
\end{equation}
where 
\begin{equation}\label{eq: KL operator}
    K_{L}=\sum_{j=0}^{L}k_{j}(x)\,\partial_{x}^{\,j}
\end{equation}
is the maximal-cut
operator~\cite{Primo:2017jtw,Frellesvig:2017aai,Bosma:2017ens}, of order $L$.

At $d=2$, the operator $K_{L}$ is the Picard--Fuchs operator of the associated
Calabi--Yau $(L-1)$-fold~\cite{Bloch:2016izu,Vanhove:2018mto,Klemm:2019dbm,Bonisch:2020qmm,Lairez:2022zkj,Vanhove:2026fth} which is defined by the vanishing locus of the second Symanzik polynomial $\mathcal{F}$ in~\eqref{eq:symanzik_poly}
\begin{align}
V(\mathcal{F}) = \left\{
\left(\sum_{i=1}^n a_i\right)\prod_{i=1}^{n}a_i\sum_{i=1}^n a_i^{-1}-x \prod_{i=1}^{n}a_i =0   
\right\} \subset \mathbb{P}^{n-1}\,.
\end{align}

The following Lemma holds~\cite{Mishnyakov:2023wpd,Mishnyakov:2023sly}
\begin{lemma}
\label{lem:em_kstructure}
Every $k_{j}$ is a polynomial in $x$. Normalising $K_{L}$ so that $k_{L}$ is monic
up to the power of $x$ in~\eqref{eq:em_kl},
\begin{align}
    k_{L}(x)&=x^{\lfloor n/2\rfloor}\prod_{k=1}^{c}\bigl(x-s_{k}\bigr) \,,
      \label{eq:em_kl}\\
    k_{L-1}(x)&=\frac{L}{2}\left(\frac{n(d-2)}{2}\,\frac{k_{L}(x)}{x}
      -(d-3)\,k_{L}'(x)\right) ,
      \label{eq:em_klm1}\\
    k_{0}(x)&=\frac{1}{2^{L}}\prod_{j=2}^{L}\bigl(2+j(2-d)\bigr)\;
      \Bigl[(4-d)\,x-(n-2)\,d-4\Bigr] \,,
      \label{eq:em_k0}
\end{align}
and for generic $d$ every coefficient satisfies
\begin{equation}\label{eq:em_orders}
    \deg k_{j}=j+1 \,,\qquad
    \operatorname{ord}_{x=0}k_{j}=\max\bigl(0,\,j-c+1\bigr) \,,
    \qquad j=0,\dots,L \,.
\end{equation}
\end{lemma}

\noindent
The pattern~\eqref{eq:em_orders} follows from
the explicit form of $K_{L}$ built from the $n$-th symmetric power of the Bessel
operator, as in~\cite{Bonisch:2020qmm,Bonisch:2021yfw,Pogel:2022vat}.

\subsection{The all-order basis}

\begin{theorem}\label{thm:em_main}
Let us choose as basis of master integrals
\begin{equation}\label{eq:em_basis}
    J_{1}=K_{L}\bullet f \,,
    \qquad
    J_{j+1}=x^{\,j-c}\;\partial_{x}^{\,j-1}\bullet f \,,
    \qquad j=1,\dots,L \,,
\end{equation}
where $f=I_{1,\dots,1}$.
Then $J=(J_{1},\dots,J_{n})^{T}$ satisfies a Pfaffian system in $d\log$ form
\begin{equation}
\label{eq:em_dlog}
    dJ=\Biggl(\;\sum_{k=0}^{c} A_{k}(d)\;d\log \beta_{k}\Biggr)J \,,
\end{equation}
with the letters~\eqref{eq:em_letters} and constant matrices $A_{k}(d)$ rational in
$d$. 
These matrices have the following structure.
\begin{enumerate}[(i)]
\item The first row of every $A_{k}$ vanishes.
\item Rows $2,\dots,L$ of $A_k$ are nonzero only for $k=0$, where they are bidiagonal,
\begin{equation}\label{eq:em_chainblock}
    (A_{0})_{j+1,j+1}=j-c \,,\qquad (A_{0})_{j+1,j+2}=1 \,,
    \qquad j=1,\dots,L-1 \,,
\end{equation}
all remaining entries of those rows being zero.
In particular, every threshold
residue $A_{k}$, $k\geq1$, has rank one, and every $A_{k}$ is carried by its last
row together with the fixed block~\eqref{eq:em_chainblock}.
\item The first column of the last row is free of $d$:
\begin{equation}\label{eq:em_col1}
    (A_{0})_{n,1}=\frac{(-1)^{c}}{(n!!)^{2}} \,,
    \qquad
    (A_{k})_{n,1}=\frac{1}{s_{k}\prod_{j\neq k}\bigl(s_{k}-s_{j}\bigr)} \,.
\end{equation}
\item The second column of the last row is
$(A_{k})_{n,2}=\operatorname*{Res}_{\beta_{k}=0}\Phi$ with
\begin{equation}\label{eq:em_col2}
    \Phi(x)=-\,\frac{x^{\,c-2}\,k_{0}(x)}{\prod_{k=1}^{c}\bigl(x-s_{k}\bigr)}\,,
\end{equation}
and $k_{0}$ as in~\eqref{eq:em_k0}; in particular $(A_{0})_{n,2}=0$ for $L\geq2$.
\item Every threshold residue has the same trace, and hence the spectrum
\begin{equation}\label{eq:em_trace}
    \operatorname{tr}A_{k}=\frac{L\,(d-3)}{2} \,,
    \qquad
    \operatorname{spec}A_{k}=\Bigl\{\,0^{\,L},\ \tfrac{L(d-3)}{2}\,\Bigr\} \,,
    \qquad k\geq1 \,,
\end{equation}
while at the origin
\begin{equation}\label{eq:em_traceA0}
    \operatorname{tr}A_{0}
    =\frac{L}{2}\Bigl(n-2c-\frac{n(d-2)}{2}+(d-3)(n-c)\Bigr) \,,
\end{equation}
and $A_{\infty}=-\sum_{k}A_{k}$ has spectrum
$\bigl\{\,1+\tfrac{j}{2}(2-d)\,\bigr\}_{j=0}^{n-1}$ modulo $\mathbb{Z}$.
\end{enumerate}
\end{theorem}

\noindent
Items (i)--(v) determine, of the $n$ entries of the last row of each residue, the
first two and the last. The remaining $\max(0,L-2)$ entries per matrix are the
values at the letters of $k_{1},\dots,k_{L-2}$. For $L=1$ and $L=2$ nothing is
left, and the whole system is fixed by~\eqref{eq:em_chainblock}--\eqref{eq:em_traceA0}.

\begin{proof}
Since $K_{L}\bullet f$ is constant, $\partial_{x}J_{1}=0$, which is (i). For
$j=1,\dots,L-1$,
\begin{equation}\label{eq:em_chain}
    \partial_{x}J_{j+1}
    =\frac{j-c}{x}\,J_{j+1}+\frac{1}{x}\,J_{j+2} \,,
\end{equation}
which is of $d\log$ type with the single letter $\beta_{0}$ and gives (ii). The chain
closes through~\eqref{eq:em_LdK}: using
$\partial_{x}^{\,L}\bullet f
=k_{L}^{-1}\bigl(J_{1}-\sum_{m=0}^{L-1}k_{m}\,\partial_{x}^{\,m}\bullet f\bigr)$
and $\partial_{x}^{\,m}\bullet f=x^{\,c-m-1}J_{m+2}$,
\begin{equation}\label{eq:em_closure}
    \partial_{x}J_{L+1}
    =\frac{L-c}{x}\,J_{L+1}
     +\frac{x^{\,L-c}}{k_{L}}\,J_{1}
     -\sum_{m=0}^{L-1}\frac{x^{\,L-m-1}k_{m}}{k_{L}}\,J_{m+2} \,.
\end{equation}

It remains to check that these coefficients have at worst simple poles at each
$\beta_{k}=0$ and vanish at least as $1/x$ at $x=\infty$. At a threshold this is
immediate from~\eqref{eq:em_kl}, whose zeros at the $s_{k}$ are simple, together
with the polynomiality of the $k_{m}$. At $x=0$, by~\eqref{eq:em_orders},
\begin{equation}
    \operatorname{ord}_{x=0}\!\left(\frac{x^{\,L-m-1}k_{m}}{k_{L}}\right)
    =L-m-1+\max(0,m-c+1)-\lfloor n/2\rfloor \;\geq\; -1 \,,
\end{equation}
with equality for $m\geq c-1$ and no pole for $m<c-1$, and likewise
$\operatorname{ord}_{x=0}(x^{L-c}/k_{L})=L-c-\lfloor n/2\rfloor=-1$. At infinity,
$\deg k_{j}=j+1$ gives
\begin{equation}
    \deg\!\left(\frac{x^{\,L-m-1}k_{m}}{k_{L}}\right)=-1 \,,
    \qquad
    \deg\!\left(\frac{x^{\,L-c}}{k_{L}}\right)=-c-1\leq -1 \,.
\end{equation}
Hence every entry of the connection is a sum of simple poles at the $\beta_{k}$,
which is~\eqref{eq:em_dlog}. Note that the bound $\lfloor n/2\rfloor\leq L$, valid
for all $L\geq1$, is what lets the order-$\lfloor n/2\rfloor$ pole of $k_{m}/k_{L}$
at $x=0$ be spread over the $L$ steps of~\eqref{eq:em_chain}, each of which absorbs
one unit.

For (iii),~\eqref{eq:em_kl} and $L-c-\lfloor n/2\rfloor=-1$ give the
$L$-independent shape
\begin{equation}\label{eq:em_col1shape}
    \frac{x^{\,L-c}}{k_{L}}=\frac{1}{x\prod_{k=1}^{c}(x-s_{k})} \,,
\end{equation}
whose residues are~\eqref{eq:em_col1}. Item (iv) is the $m=0$ term of~\eqref{eq:em_closure}, with $L-1-\lfloor n/2\rfloor=c-2$.

For (v), the connection has trace
\begin{equation}\label{eq:em_trA}
    \operatorname{tr}A(x)=\sum_{j=1}^{L}\frac{j-c}{x}
      -\frac{k_{L-1}(x)}{k_{L}(x)} \,,
\end{equation}
so $\operatorname{tr}A_{k}$ is minus the residue of $k_{L-1}/k_{L}$ at
$\beta_{k}=0$, which~\eqref{eq:em_klm1} evaluates. Since $k_{L}$ has a simple zero
at $s_{k}\neq0$, the quotient $k_{L}'/k_{L}$ has residue $1$ there while $k_{L}/x$
is regular, giving~\eqref{eq:em_trace}; rank one then fixes the spectrum. At the
origin $k_{L}'/k_{L}$ has residue $\operatorname{ord}_{x=0}k_{L}=n-c$, and
$\sum_{j=1}^{L}(j-c)=\tfrac{L}{2}(n-2c)$, which is~\eqref{eq:em_traceA0}. The
behaviour at infinity follows from the $\theta$-form of $\mathcal{L}_{n}$, whose
leading block is $\prod_{j=0}^{n-1}\bigl(\theta+1+\tfrac{j}{2}(2-d)\bigr)$.
\end{proof}

\begin{example}
The bases~\eqref{eq:em_basis} for the first five loop orders are the following.

\begin{center}
\begin{tabular}{@{}cccl@{}}
\toprule
$L$ & $c$ & letters & basis $J^{T}$ \\
\midrule
$1$ & $1$ & $x,\;x-4$               & $\bigl(K_{1}\bullet f,\;f\bigr)$ \\
$2$ & $2$ & $x,\;x-9,\;x-1$         & $\bigl(K_{2}\bullet f,\;f/x,\;f'\bigr)$ \\
$3$ & $2$ & $x,\;x-16,\;x-4$        & $\bigl(K_{3}\bullet f,\;f/x,\;f',\;xf''\bigr)$ \\
$4$ & $3$ & $x,\;x-25,\;x-9,\;x-1$  & $\bigl(K_{4}\bullet f,\;f/x^{2},\;f'/x,\;f'',\;xf'''\bigr)$ \\
$5$ & $3$ & $x,\;x-36,\;x-16,\;x-4$ & $\bigl(K_{5}\bullet f,\;f/x^{2},\;f'/x,\;f'',\;xf''',\;x^{2}f''''\bigr)$ \\
\bottomrule
\end{tabular}
\end{center}

\noindent
The residue matrices for $L\leq3$ are as follows; for $L=4,5$ they are obtained
in the same way from $K_{L}$. For $L=1$, with $k_{1}=x(x-4)$ and
$k_{0}=\tfrac12\bigl[(4-d)x-4\bigr]$,
\begin{equation}
    A_{0}=\begin{pmatrix} 0 & 0\\[2pt] -\tfrac14 & -\tfrac12 \end{pmatrix} \,,
    \qquad
    A_{1}=\begin{pmatrix} 0 & 0\\[2pt] \tfrac14 & \tfrac12(d-3) \end{pmatrix} \,.
\end{equation}
For $L=2$,
\begin{equation}
    A_{0}=\begin{pmatrix}
       0 & 0 & 0\\[2pt]
       0 & -1 & 1\\[2pt]
       \tfrac19 & 0 & -\tfrac d2
    \end{pmatrix} \,,
\end{equation}
\begin{equation}
    A_{1}=\begin{pmatrix}
       0 & 0 & 0\\[2pt]
       0 & 0 & 0\\[2pt]
       \tfrac1{72} & -\tfrac18(d-3)(5d-16) & d-3
    \end{pmatrix} \,,
    \qquad
    A_{2}=\begin{pmatrix}
       0 & 0 & 0\\[2pt]
       0 & 0 & 0\\[2pt]
       -\tfrac18 & \tfrac18\,d\,(d-3) & d-3
    \end{pmatrix} \,.
\end{equation}
For $L=3$,
\begin{equation}
    A_{0}=\begin{pmatrix}
       0 & 0 & 0 & 0\\[2pt]
       0 & -1 & 1 & 0\\[2pt]
       0 & 0 & 0 & 1\\[2pt]
       \tfrac1{64} & 0 & \tfrac14 d(d-4) & -2
    \end{pmatrix} \,,
\end{equation}
\begin{equation}
    A_{1}=\begin{pmatrix}
       0 & 0 & 0 & 0\\[2pt]
       0 & 0 & 0 & 0\\[2pt]
       0 & 0 & 0 & 0\\[2pt]
       \tfrac1{192} & \tfrac18(d-3)(3d-10)(3d-8) & -(d-3)(3d-10) & \tfrac32(d-3)
    \end{pmatrix} \,,
\end{equation}
\begin{equation}
    A_{2}=\begin{pmatrix}
       0 & 0 & 0 & 0\\[2pt]
       0 & 0 & 0 & 0\\[2pt]
       0 & 0 & 0 & 0\\[2pt]
       -\tfrac1{48} & -\tfrac18(d-3)(d-2)(3d-8) & 2(d-3) & \tfrac32(d-3)
    \end{pmatrix} \,.
\end{equation}

\end{example}

\section*{Acknowledgments}
We would like to thank Vsevolod Chestnov and Pierpaolo Mastrolia for valuable discussions and collaboration on early stages of the project.
W.F. was supported by the Polish National Science Centre (NCN) under grant 2023/50/A/ST2/00224.
Y.G. was supported by JSPS KAKENHI Grant Number JP24K06680.
S.-J. M.-H. was supported by JSPS KAKENHI Grant Number JP26K06838.
N.T. was supported by JSPS KAKENHI Grant Number JP26K06844.
G.B.'s research is supported by the Italian MIUR under 
contract 20223ANFHR (PRIN2022), 
by the ERC (NOTIMEFORCOSMO, 101126304),
and by the INFN initiatives \textit{Amplitudes} and \textit{TPPC}.

\medskip

\noindent
{\bf AI disclosure}
During the preparation of this manuscript, AI tools were used to simulate a pre-referee review process.
These AI tools were used to improve the overall exposition, enhance clarity, and identify potential inconsistencies and oversights in early drafts. After using these tools, the authors thoroughly reviewed and edited the manuscript, and take full responsibility for the final content, accuracy, and conclusions of this article.

{\footnotesize
\bibliographystyle{abbrv}
\bibliography{refs.bib}
}

\bigskip

\noindent
\begin{tabular}{l}
    \textbf{Giacomo Brunello} \\
    Scuola Normale Superiore, Piazza dei Cavalieri 7, 56126, Pisa, Italy, \\
    INFN Sezione di Pisa, Largo Pontecorvo 3, 56127 Pisa, Italy\\
    \texttt{giacomo.brunello@sns.it}
\end{tabular}

\bigskip
\noindent
\begin{tabular}{l}
    \textbf{Wojciech Flieger} \\
    Faculty of Physics, Astronomy and Applied Computer Science, Jagiellonian University \\
    ul. \L{}ojasiewicza 11, 30-348 Krak\'{o}w, Poland \\
    \texttt{wojciech.flieger@uj.edu.pl}
\end{tabular}

\bigskip
\noindent
\begin{tabular}{l}
    \textbf{Yoshiaki Goto} \\
    General Education, Otaru University of Commerce\\
    3-5-21, Midori, Otaru 047-8501, Japan\\
    \texttt{goto@res.otaru-uc.ac.jp}
\end{tabular}

\bigskip
\noindent
\begin{tabular}{l}
    \textbf{Saiei-Jaeyeong Matsubara-Heo} \\
    Graduate School of Information Sciences, Tohoku University\\
    6-3-09 Aramaki-Aza-Aoba, Aoba-ku,     Sendai 980-8579, Japan \\
    \texttt{saiei@tohoku.ac.jp}
\end{tabular}

\bigskip
\noindent
\begin{tabular}{l}
    \textbf{Nobuki Takayama} \\
    Department of Mathematics, Kobe University, Kobe, 657-8501, Japan \\
    \texttt{takayama@math.kobe-u.ac.jp}
\end{tabular}

\end{document}